\documentclass[namedreferences]{SolarPhysics}
\usepackage[optionalrh,solaenum]{spr-sola-addons} 
\usepackage{savesym}
\savesymbol{iint}
\savesymbol{iiint}

\usepackage{graphicx}                    
\usepackage{amsmath}
\usepackage{color}                       
\usepackage{txfonts}
\usepackage{url}                         

\newcommand{\Sunrise}{\mbox{\textit{Sunrise} }}

\newcommand{\aap}{    {\it Astron. Astrophys.}}

\newcommand{\apj}{    {\it Astrophys. J.}}

\newcommand{\solphys}{{\it Solar Phys.}}

\begin{document}

\begin{article}

\begin{opening}

\title{The Imaging Magnetograph eXperiment (IMaX) for the \Sunrise balloon-borne 
solar observatory}

%
   \author{V.~\surname{Mart\'\i nez Pillet},$^{1}$\sep
          J.C.~\surname{del Toro Iniesta}$^{2}$\sep 
	  A.~\surname{\'Alvarez-Herrero}$^{3}$\sep 
	  V.~\surname{Domingo}$^{4}$\sep 
	  J. A.~\surname{Bonet}$^{1}$\sep 
	  L.~\surname{Gonz\'alez Fern\'andez}$^{3}$\sep 
	  A.~\surname{L\'opez Jim\'enez}$^{2}$\sep 
	  C.~\surname{Pastor}$^{3}$\sep 
	  J.L.~\surname{Gasent Blesa}$^{4}$\sep
	  P.~\surname{Mellado}$^{2}$\sep 
	  J.~\surname{Piqueras}$^{5}$\sep
	  B.~\surname{Aparicio}$^{2}$\sep
	  M.~\surname{Balaguer}$^{2}$\sep 
	  E.~\surname{Ballesteros}$^{1}$\sep
	  T.~\surname{Belenguer}$^{3}$\sep 
	  L.R.~\surname{Bellot Rubio}$^{2}$\sep
	  T.~\surname{Berkefeld}$^{6}$\sep
	  M.~\surname{Collados}$^{1}$\sep
	  W.~\surname{Deutsch}$^{5}$\sep
	  A.~\surname{Feller}$^{5}$\sep
	  F.~\surname{Girela}$^{2}$\sep
	  B.~\surname{Grauf}$^{5}$\sep
	  R.L.~\surname{Heredero}$^{3}$\sep
	  M.~\surname{Herranz}$^{2}$\sep
	  J.M.~\surname{Jer\'onimo}$^{2}$\sep
	  H.~\surname{Laguna}$^{3}$\sep
	  R.~\surname{Meller}$^{5}$\sep
	  M.~\surname{Men\'endez}$^{3}$\sep 
	  R.~\surname{Morales}$^{2}$\sep 
	  D.~\surname{Orozco Su\'arez}$^{2}$\sep 
	  G.~\surname{Ramos}$^{3}$\sep
	  M.~\surname{Reina}$^{3}$\sep 
	  J.L.~\surname{Ramos}$^{2}$\sep
	  P.~\surname{Rodr\'\i guez}$^{4}$\sep
	  A.~\surname{S\'anchez}$^{3}$\sep
	  N.~\surname{Uribe-Patarroyo}$^{3}$\sep
	  P.~\surname{Barthol}$^{5}$\sep
	  A.~\surname{Gandorfer}$^{5}$\sep
	  M.~\surname{Knoelker}$^{7}$\sep
	  W.~\surname{Schmidt}$^{6}$\sep
	  S.K.~\surname{Solanki}$^{5}$\sep 
	  S.~\surname{Vargas Dom\'\i nguez}$^{1}$
          }

%
\runningauthor{Mart\'\i nez Pillet {\it et al.}}
\runningtitle{The Imaging Magnetograph eXperiment of \Sunrise}

%
  \institute{$^{1}$ Instituto de Astrof\'\i sica de Canarias, E-38200, La Laguna, 
  		    Tenerife, Spain
                    email: \url{vmp@iac.es} \\ 
             $^{2}$ Instituto de Astrof\'\i sica de Andaluc\'\i a (CSIC), Apdo. de Correos 3004, E-18080, Granada, Spain \\
             $^{3}$ Instituto Nacional de T\'ecnica Aeroespacial, E-28850, Torrej\'on de Ardoz, Madrid, Spain \\
             $^{4}$ Grupo de Astronom\'\i a y Ciencias del Espacio (Univ. de Valencia), E-46980, Paterna, Valencia, Spain \\
             $^{5}$ Max-Planck-Institut f\"ur Sonnensystemforschung, 37191, Katlenburg-Lindau, Germany \\
             $^{6}$ Kiepenheuer-Institut f\"ur Sonnenphysik, 79104, Freiburg, Germany \\
             $^{7}$ High Altitude Observatory (NCAR), 80307-3000, Boulder, USA 
             }

\begin{abstract}
The Imaging Magnetograph eXperiment (IMaX) is a spectropolarimeter built by
four institutions in Spain that flew on board the \Sunrise balloon-borne
telesocope in June 2009 for almost six days over the Arctic Circle. As a
polarimeter IMaX uses fast polarization modulation (based on the use of two
liquid crystal retarders), real-time image accumulation, and dual beam
polarimetry to reach polarization sensitivities of 0.1 \%. As a spectrograph,
the instrument uses a LiNbO$_3$ etalon in double pass and a narrow band
pre-filter to achieve a spectral resolution of 85 m\AA.  IMaX uses the high
Zeeman sensitive line of Fe {\sc i} at 5250.2 \AA~and observes all four Stokes
parameters at various points inside the spectral line. This allows vector
magnetograms, Dopplergrams, and intensity frames to be produced that, after
reconstruction, reach spatial resolutions in the 0.15-0.18
arcsec range over a 50$\times$50 arcsec FOV. Time cadences vary between ten
and 33 seconds, although the shortest one only includes longitudinal
polarimetry. The spectral line is sampled in various ways depending on the
applied observing mode, from just two points inside the line to 11 of them. All
observing modes include one extra wavelength point in the nearby continuum.
Gauss equivalent sensitivities are four Gauss for longitudinal fields and 80 Gauss for
transverse fields per wavelength sample. The LOS velocities are estimated with
statistical errors of the order of 5-40 m s$^{-1}$.  The design, calibration
and integration phases of the instrument, together with the implemented
data reduction scheme are described in some detail.
\end{abstract}

%
%
\end{opening}

\section{Introduction}

Increasing our understanding of solar magnetism poses high demands on spatial
resolution and polarimetric sensitivity. One-meter aperture telescopes allow 
solar structures with typical sizes of about 100 km (in the visible) to be resolved
which has proven very beneficial in identifying fundamental ingredients of 
active region plages, penumbral filaments, umbral dots and solar granulation
(Scharmer, \citeyear{Scharmer09}; Lites {\it et al.}, \citeyear{Lites04}). The requirements for
the detectability of the (often weak) magnetic fields that interact with the
plasma at the solar surface are very  stringent. Recent
magnetoconvection simulations of the conditions in the quiet Sun show
distributions of field strengths that increase monotonically from the kG fields
to values of about one Gauss (see Pietarila Graham {\it et al.}, \citeyear{Pietarila09}). One
Gauss of longitudinal field produces a polarization signal that requires the
detection of 4800$^2\approx 2\cdot 10^7$ photoelectrons in a time scale short
enough for this field not to evolve considerably (seconds). This is a very
demanding measurement. For transverse fields of a similar magnitude, the
situation is much more discouraging as one needs now to detect 2500$^4\approx
4\cdot 10^{13}$ photoelectrons (the rationale for these numbers is given in
Section\ \ref{sec:sample}). It is in this context that a number of solar physics
projects have been conceived in the last decade for the construction of
large-aperture telescopes and high-sensitivity polarimeters, both ground-based
(4 m class telescopes; see Rimmele {\it et al.}, \citeyear{Rimmele09} and Collados,
\citeyear{Collados09}) and space-borne ({\em Hinode}, Kosugi, \citeyear{Kosugi07}). Led
by the Max Planck Institut f\"ur Sonnensystemforschung (MPS,
Katlenburg-Lindau, Germany), the \Sunrise project (Barthol {\it et al.},
\citeyear{Barthol10}) is a multi-national effort of that type aimed at imaging the
near-UV continuum and at observing the photospheric magnetic field and velocity
from a balloon platform with a one m aperture telescope. To this end \Sunrise 
uses the Imaging Magnetograph eXperiment (IMaX) in order to map the
line-of-sight (LOS) velocity and the complete magnetic field vector using
spectropolarimetric observations of a photospheric line.

   \begin{figure*}
   \centering
   \includegraphics[width = 300pt]{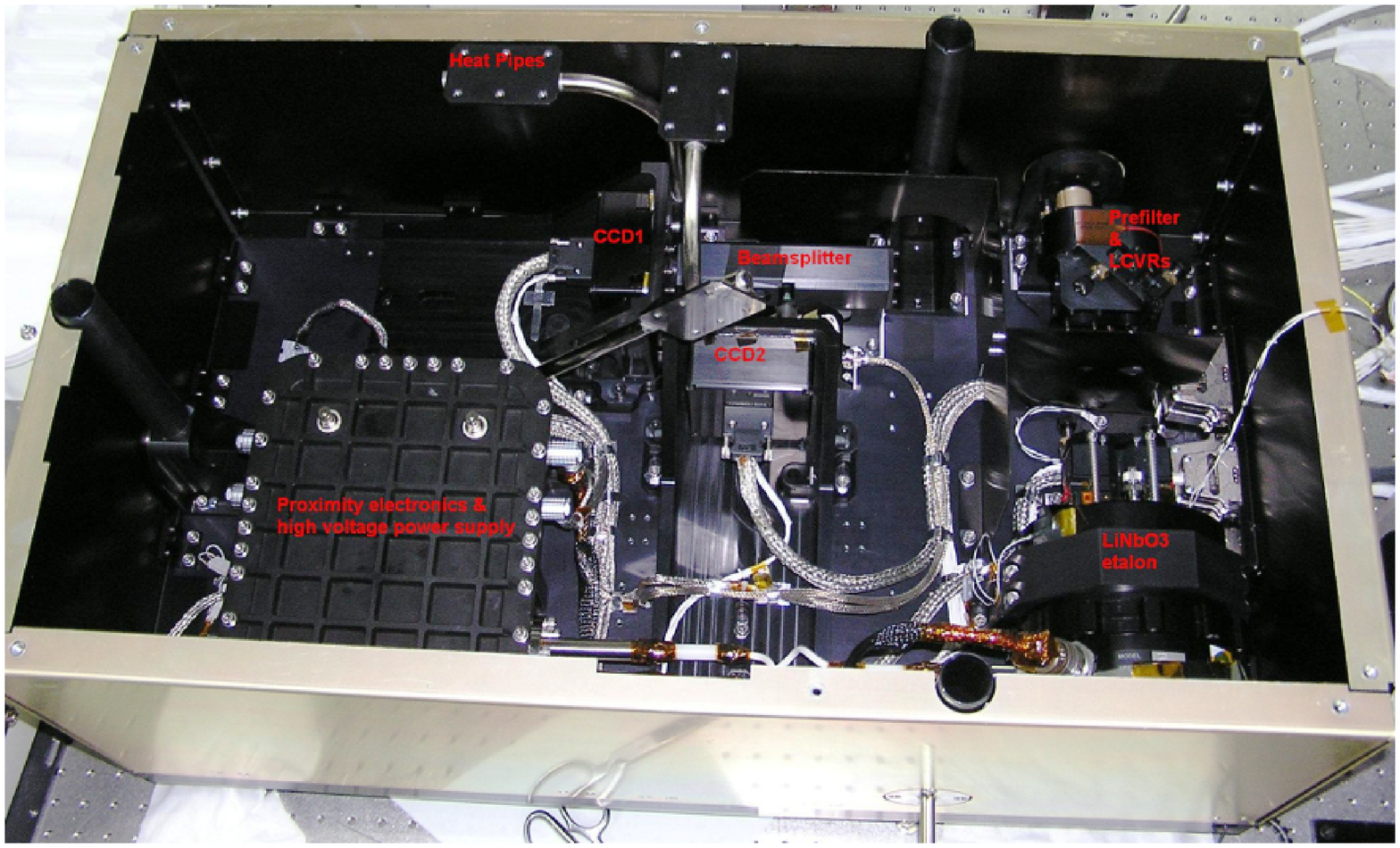}
   \includegraphics[width = 300pt]{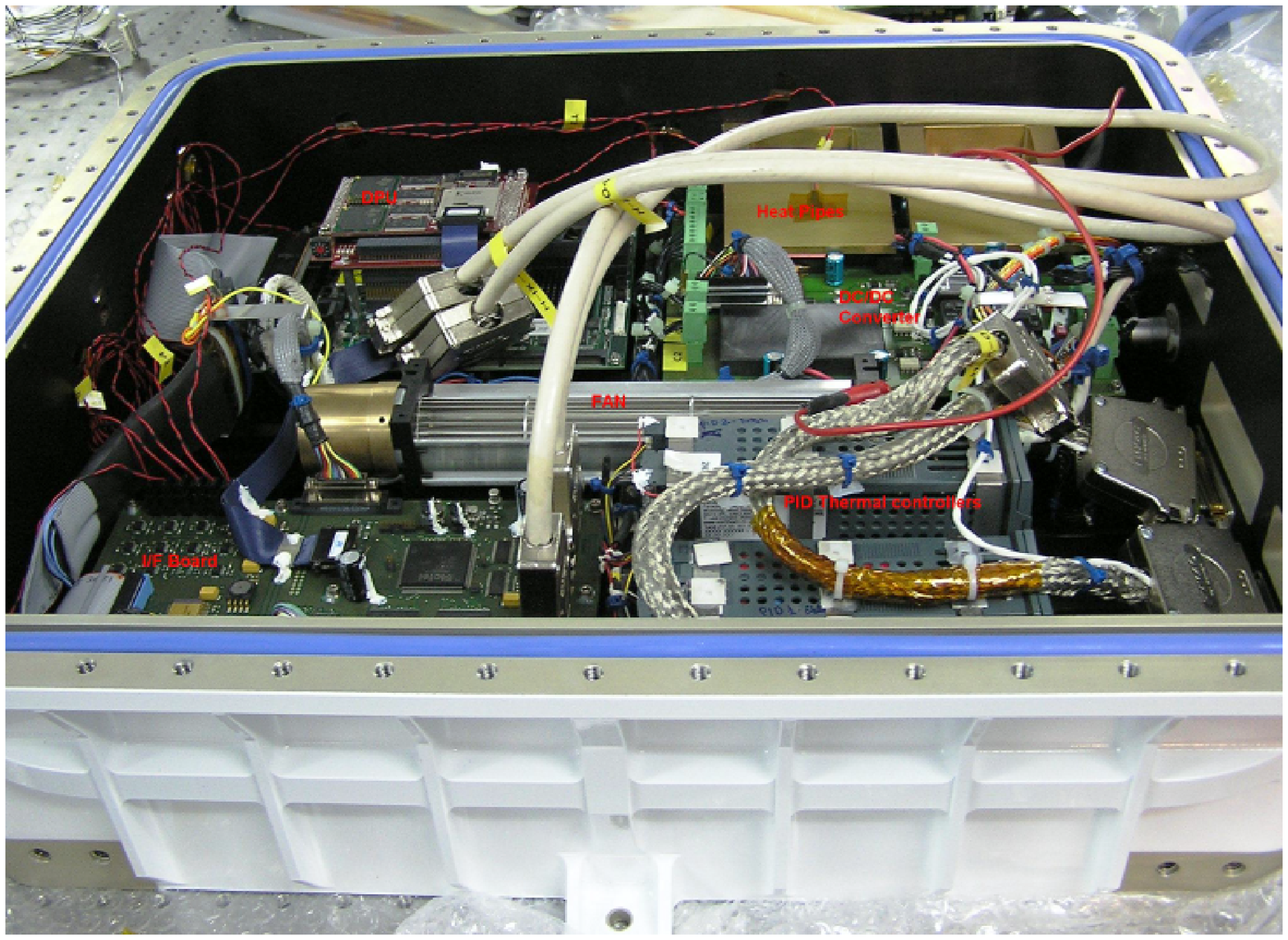}
   \caption{Top: IMaX optical enclosure just after integration and
   verification at INTA.  The top cover (a radiator) is not in place yet.
   Labels correspond to the pre-filter and LCVR mounting, the LiNbO$_3$ etalon
   enclosure, the polarizing beamsplitter, the CCDs, the heat pipes, and the
   proximity electronics enclosure in the lower left. Bottom: IMaX main
   electronic enclosure. Labels correspond to the control computer and real
   time electronics (DPU), the DC/DC converters, the heat pipes, the fan, the interface
   board, and the thermal controllers.
               }
              \label{fig:IMaX}%
    \end{figure*}

IMaX has been designed, built, and calibrated by a consortium of four
institutions in Spain. The Instituto de Astrof\'\i sica de Canarias (IAC,
Tenerife) is the leading institution and has been in charge of the conceptual
design, management of the consortium, and the ground-based control software.
The Instituto de Astrof\'\i sica de Andaluc\'\i a (IAA -- CSIC, Granada) has
been in charge of all the electronic aspects of the instrument. These include
the detectors, real time electronics, control computer and software, power
supply and distribution, and harness. The Instituto Nacional de T\'ecnica
Aeroespacial (INTA, Torrej\'on de Ardoz) has developed the optical definition
of the instrument, its opto-mechanical concept, and the thermal sub-system. The
Grupo de Astronom\'\i a y Ciencias del Espacio (GACE, Valencia) has provided
the three enclosures (two of them pressurized) used by the instrument. The
Assembly, Integration and Verification (AIV) of the instrument (and
various subsystems) has been
performed at INTA facilities prior to the submission of the instrument to
MPS.  A conceptual description of the instrument has already been provided
elsewhere (Mart\'\i nez Pillet, \citeyear{Marpillet04}). The description presented
there is still valid with the exception of the spectral line used by the
instrument at that time, 
the detectors, and the pre-filter bandpass. The status of the
project at critical design stage can be found in \'Alvarez-Herrero {\it et al.},
(\citeyear{Alvarez06a}). An illustration of the instrument can be seen in Figure\
\ref{fig:IMaX}. The top panel shows the Optical Bench Enclosure (OBE) and the bottom
one is a picture of the Main Electronics Enclosure (MEE). The black anodized
box inside the OBE is the Proximity Electronics Enclosure (PEE).

This paper describes the IMaX instrument in detail, from the conceptual level
to the complete calibration process at INTA facilities, at MPS facilities, and
in the launch field of the ESRANGE Space Center (Kiruna, Sweden). IMaX data
reduction and calibration is also covered in some length. The paper is
intended to contain all the information needed for analyzing the data obtained by
the instrument during the June 2009 flight.

\section{Instrument Concept}
\label{sec:introduction}

\subsection{Instrument Requirements}
\label{sec:requirements}

To accomplish its scientific objectives, IMaX must perform near
diffraction-limited imaging with high spectroscopic resolving power and very
high sensitivity polarimetry. Differential imaging in selected wavelengths of a
Zeeman-sensititve spectral line are needed to extract the necessary information
to infer the vector magnetic field and the LOS velocity of the solar plasma.
These three characteristics should be borne in mind all along the design,
development, and construction phases of the instrument. A great deal of
prior studies had to be carried out in order to select the minimum
performance of the instrument and a significant number of trade-offs have been
taken into account. 

First of all, IMaX had to preserve the imaging capabilities of the
\Sunrise one-meter telescope to the maximum. 
The system-level specification for the image
quality of IMaX was set independently of the rest of the optical setup in front
of the instrument (ISLiD and telescope, Gandorfer {\it et al.}, \citeyear{Gandorfer10}).
\footnote{ISLiD is an acronym for Image
Stabilization and Light Distribution system.} It has been given in terms of the
Strehl ratio that the instrument should have for an aberration free image at
F4 (the nominal entrance focus of IMaX and the interface with the ISLiD). 
It requires that the Strehl ratio in the IMaX image plane should be better
than 0.9 for nominal optical configuration and that the effect of fabrication
and alignment tolerances should not degrade the Strehl ratio to less than 0.8
($\lambda$/14) over the entire FOV. 
This requirement was set after confirmation of the high optical quality
achievable by the etalon as stated by the manufacturer.
Several factors can deteriorate the image quality such as
internal aberrations due to misalignments or poor optical quality of the IMaX
components and errors induced from residual jittering of the platform among
others. To guarantee a good recovery of near-diffraction images, we decided to
use a phase diversity mechanism in order to evaluate the point spread function
(PSF) of the instrument right after every observing run, so that the observed
images can be deconvolved from this PSF. Extensive calculations (Vargas Dom\'\i
nguez, \citeyear{Vargas09}) led us to the conclusion that an rms wavefront error
(WFE) of $\lambda/5$ can be accepted for IMaX in order not to lose spatial
resolution after PSF deconvolution.

Second, the spectroscopic and polarimetric capabilities of the instrument,
along with the sensitivity of the finally selected spectral line, are
intimately involved to ensure the instrument magnetic and kinematic
performance. The community has reached a general consensus that
a minimum signal-to-noise ratio (S/N) of $10^3$ is required to detect the
weak polarization signals present in the internetwork part of 
the quiet Sun (even though higher values are desirable). S/N is the key
parameter governing the photometric and polarimetric sensitivity of the
instrument since it yields the minimum intensity, $\delta I$, and 
minimum polarization degree, $\delta p$ with $p$ the polarization
degree itself, that can be detected:
\footnote{Note here that S/N is defined as the ratio between signal at a given
wavelength over noise at the same wavelength, whereas the common use of this figure
of merit is computed at continuum wavelengths.}
\begin{equation}
\label{eq:s/n}
\frac{\delta I}{I} = \frac{1}{S/N};~~~~~~~~~~\frac{\delta p}{p} \geq \frac{\sqrt{1+3/p^2}}{S/N}.
\end{equation}
Polarimetric accuracy is usually measured through the so-called polarimetric
efficiencies (see Section\ \ref{sec:polarimetry}). Minimum IMaX efficiencies for
Stokes $Q$, $U$, and $V$ were set to 0.45.  A number of numerical experiments,
including synthetic observations built with magnetohydrodynamical simulations
(V\" ogler {\it et al.}, \citeyear{Vogler05})
of the solar photosphere and analyzed with inversions of the radiative transfer
equation, were carried out (Orozco Su\'arez, 2008; Orozco Su\'arez {\it et al.},
2010). These experiments indicated that a spectral resolution of the order of
50 -- 100 m\AA\ (with a preference for the high resolution range), in four
wavelengths  across the line profile plus one in the continuum, is enough to
achieve accurate values of the vector magnetic field and the LOS velocity from
post-processing of the IMaX data.

\subsection{Basic Design Concepts for a High Sensitivity Polarimeter}
\label{sec:general}

As mentioned in the introduction, the polarimetric sensitivity required to measure one Gauss
of longitudinal field requires typically 10$^7$ photoelectrons or so.
Currently available CCDs have full wells of about 10$^5$ electrons and in a typical exposure
one can only fill a fraction of that. This readily tells us that with single exposures 
it is impossible to reach these magnetic sensitivities. It is for this reason 
that modern solar polarimeters introduced image accumulation to add a number, 
$N_{\rm A}$, of single exposures to achieve the desired sensitivity or S/N.
This scheme was successfully employed and popularized by the Advanced
Stokes Polarimeter (ASP, Elmore, \citeyear{Elmore92}). Other polarimeters that follow
this scheme of real-time addition of exposures to achieve a $S/N\approx 10^3$ are the 
Tenerife Infrared Polarimeter (TIP I and II,
Mart\'\i nez Pillet {\it et al.}, \citeyear{Marpillet99}; Collados {\it et al.},
\citeyear{Collados07});
ZIMPOL (Gandorfer {\it et al.}, \citeyear{Gandorfer04} and references therein) and the
{\em Hinode}/SP (Lites, \citeyear{Lites01}). Accumulation, but 
after applying reconstruction techniques (speckle or blind deconvolution), is becoming 
also popular in ground-based filter magnetographs (Bello Gonz\'alez and Kneer, 
\citeyear{Bello08};
Viticchi\'e {\it et al.}, \citeyear{Viticchie09}; van Noort and Rouppe van der Voort, 
\citeyear{Vannoort08}).

An additional complication results from the fact that a polarimeter is a
differential imager (see Mart\'\i nez Pillet, \citeyear{Marpillet06}). The polarization
signals are inferred from the subtraction of two (or more) images, 
referred to as the polarization modulation images, taken at different times. 
Any difference between them that originated from atmospheric seeing, pointing
errors or even solar evolution is interpreted as a real polarization signal.
There are two complementary ways to minimize this effect, as explained, {\it e.g.},
by Del Toro Iniesta (2003). First, one implements fast temporal polarization
modulation so that the time between the images that will be subsequently
subtracted is minimized. This imposes serious requirements on the polarization
modulator that is normally a spinning retarder (like in the ASP or in {\em
Hinode}) or an electro-optical retarder based on some type of liquid crystal
material (typically ferroelectric as in TIP or nematic as in IMaX). Slower
polarization modulators with step motors (such as in SOHO/MDI, Scherrer,
\citeyear{Scherrer95}) are only viable options at low spatial resolution and with
accurate pointing in the absence of atmospheric disturbances. The temporal
polarization modulation is then synchronized to the detector readout which
typically requires careful selection of the CCD chip and its shutter concept.
Second, to minimize jittering-induced spurious signals one measures the
various polarization modulation images in the two orthogonal linear states
produced by a polarizing beamsplitter (as originally
proposed by Lites, \citeyear{Lites87}; sometimes also referred as a dual-beam
configuration). By combining the images obtained in these two orthogonal
states, one not only makes use of all the available photons, but also cancels
out, to first order, the fluctuations produced by Stokes $I$ to the other
Stokes parameters due to image motion or solar evolution.  The combination of
these three properties, image accumulation to increase S/N, rapid
polarization modulation, and dual-beam configuration is at the core of all
modern high sensitivity polarimeters (such as the {\em Hinode}/SP instrument).
IMaX also uses these three properties.  While in the case of {\em Hinode}/SP
the pointing of the platform and the excellent performance of the on-board
correlation tracker (Shimizu {\it et al.},  \citeyear{Shimizu08}) have made the use of the
two orthogonal states measured by the SP less relevant (Lites, private
communication), the expected pointing errors from a balloon gondola suggested
its implementation. Indeed, early in the project we used the pointing errors
published by the Flare Genesis Experiment (FGE, Bernasconi {\it et al.},
\citeyear{Bernasconi00}) to simulate the spurious polarization signals introduced
by them. More specifically, a single camera with an image jittering of five arcsec
rms and a white noise spectrum for frequencies up to 10 Hz was used (see the
Figure 3 in Bernasconi {\it et al.}, \citeyear{Bernasconi00}). 
The residual motions were partially compensated by a standard
attenuation curve of a typical correlation tracker. An image of the granulation
was then subjected to such residual jitter and the differences between the
consecutive displaced frames analyzed. This test showed that the expected level
of spurious polarization signal was $10^{-2}$, an order of magnitude larger
than the intended polarimetric sensitivity, hence proving the need for a
dual-beam polarimeter in IMaX.

The performance of the instrument during its 2009 flight is listed in Table \ref{tab:perf}.

   \begin{table}
      \caption[]{IMaX requirements and performance}
         \label{tab:perf}
         \vspace{-0.4cm}
     $$ 
         \begin{array}{p{0.5\linewidth}l}
            \hline
            \noalign{\smallskip}
             Wavelength & {\rm 5250.2~\AA}  \\
             Spectral resolution $\delta\lambda$& {\rm 65~m\AA~(85~m\AA~ Gaussian)}  \\
             Spatial sampling & {\rm 0.055~arcsec/px}  \\
             Focal depth at F4& \pm {\rm 0.66~mm}  \\
             Lateral magnification & {\rm 1.8}  \\
             Image quality (only IMaX) & {\rm Strehl}~0.7~(\lambda/11)  \\
             Image quality (in flight) & {\rm Strehl}~0.3~\lambda/5.4  \\
             $N_\lambda$ (number of wavelengths)& {\rm 2,4,11 + continuum}  \\
             $N_{\rm p}$ (number of Stokes parameters)& 2,4  \\
             $N_{\rm A}$ (number of image accumulations)& {\rm variable,~6~for}~N_{\rm p}=4,~N_\lambda=5  \\
	     $t_{\rm ex}$ (individual exposures) & {\rm 246~ms} \\
	     Field of view & \approx 50\times 50{\rm~arcsec}\\
	     Wavelength drift over FOV & \approx 28 {\rm~m\AA }\\
	     $(S/N)_i$ (photon noise)& {\rm 850-1000~(5~G}~B_{\rm L},~{\rm 80~G}~B_{\rm T})\\
	     Exp. time for $N_\lambda$=5, $N_{\rm p}$=4, $N_{\rm A}$=6 &{\rm  33~s}\\
	     Exp. time for $N_\lambda$=1, $N_{\rm p}$=4, $N_{\rm A}$=6 &{\rm  6~s}\\
            \noalign{\smallskip}
            \hline
            \noalign{\smallskip}
            OBE weight &~49.0~{\rm kg}   \\
            EE weight &~26.7~{\rm kg}    \\
            Power  &~191/234~{\rm W~(mean/peak)}    \\
            Data rate  &~996~{\rm kB~s}^{-1}    \\
            \noalign{\smallskip}
            \hline
         \end{array}
     $$ 
   \end{table}
%

\subsection{Functional Concept and Trade-offs}
\label{sec:trade-offs}

Solar magnetographs must be capable of doing polarimetry, spectroscopy, and
imaging with the highest possible quality. These three different capabilities
make them necessarily complex instruments. IMaX selected the well-known
spectral line of Fe {\sc i} at 5250.2 \AA~($g=3$, $\chi_{\rm i}=0.121$ eV) as being 
the highest Zeeman sensitivity line in the green part of the spectrum. 

\subsubsection{Polarimetry}
\label{sec:polarimetry}

In the different observing modes of the instrument, a single exposure shot
corresponds to one of the polarization modulation states out of the $N_{\rm p}$ being
used by the instrument in this mode and to one specific wavelength out of the
$N_\lambda$ being observed. Of these individual exposures, IMaX real-time
electronics adds $N_{\rm A}$ of them to produce one single final image. If the
exposure time of a single shot is $t_{\rm ex}$, then the total time used to
observe all polarization states in all wavelength points is clearly $N_{\rm A} N_{\rm p}
N_\lambda t_{\rm ex}$. The observations made at continuum wavelength have
an effective exposure time of $N_{\rm A} N_{\rm p} t_{\rm ex}$. 

For the polarization analysis, IMaX uses two custom nematic liquid crystal
variable retarders (LCVRs) 
and a commercial polarizing beamsplitter. The optical axis of the first
LCVR (as seen by the light entering IMaX) is oriented parallel to one of the
linear polarization directions of the beamsplitter. The second LCVR's
optical axis is oriented at
45$^\circ$ relative to this direction. Polarization modulation is achieved by
driving the LCVRs with specific sets of voltages.  This combination of two
LCVRs is very efficient in producing, either four linearly independent
modulation states $[I_1,I_2,I_3,I_4]$ ($N_{\rm p}=4$) for vector polarimetry or two
modulation states providing the canonical combination of $I+V$ and $I-V$ for
longitudinal polarimetry ($N_{\rm p}=2$). This flexibility is very specific of LCVRs
and cannot be found in modulators based on rotating waveplates or on
ferroelectric liquid crystals (Mart\'\i nez Pillet, \citeyear{Marpillet04}).  In
the case of vector polarimetry, the modulation states are related to the Stokes
vector ($[S_1,S_2,S_3,S_4]=[I,Q,U,V]$) by the so called modulation matrix 
${\bf M}$:
\begin{equation}
     I_i = \sum_{j=1}^{4} M_{ij}S_j,~~~~i=1,2,3,4,
\end{equation}
with the demodulation matrix (that needed to deduce the Stokes parameters from
the measurements) being simply ${\bf D}={\bf M^{-1}}$. The coefficients of
${\bf D}$ are used in turn to estimate the so-called polarimetric efficiencies 
\begin{equation}
     \epsilon_i = \left( N_{p}\sum_{j=1}^{4} D_{ij}^2 \right)^{-1/2},~~~~i=1,2,3,4.
\end{equation}
If a modulation scheme weights $Q$, $U$, and $V$ equally, and a combination of
LCVR voltages is found that maximizes the S/N, then $\epsilon_{2,3,4}$ are
known to reach a maximum value of $1/\sqrt{3}$. $\epsilon_{1,{\rm max}}=1$. The
derivation of these properties is conveniently explained in Del Toro Iniesta
and Collados (\citeyear{Deltoro00}).  With these definitions, the signal-to-noise
ratio of the accumulated image in each Stokes parameter $(S/N)_i$ from each
individual camera can be derived to be:
\begin{equation}
\label{eq:numaccu}
(S/N)_i=(s/n)\,\epsilon_i \sqrt{N_{p}N_{\rm A}},~~~~i=1,2,3,4,
\end{equation}
with $(s/n)$ being the signal-to-noise ratio of the individual exposures
(integrated for $t_{\rm ex}$ seconds; Mart\'\i nez Pillet  \citeyear{Marpillet99}).

The polarization analysis is made by a polarizing beamsplitter near the CCD cameras.
In between the LCVRs and the beamsplitter  there are several optical elements, including
lenses, three folding mirrors, and the etalon. While the LCVRs change state, these 
optical elements remain constant and thus induce no variation of the polarimetric efficiencies
compared to the case when the beamsplitter is right after the LCVRs. However, the actual
coefficients of the modulation matrix ${\bf M}$ do depend on the actual polarization
properties of these optical elements.

\subsubsection{Spectroscopy}
\label{sec:spectroscopy}

For the spectral analysis, IMaX uses a combination of two optical systems: a
one\break \AA~FWHM pre-filter and a double-pass LiNbO$_3$ etalon. While these etalons
have already been used in the past in solar astronomy (in the FGE balloon in
particular, Bernasconi {\it et al.}, \citeyear{Bernasconi00}), the concept used in IMaX is
novel.  The rationale of the IMaX spectroscopic concept is as follows.
According to Section\ \ref{sec:requirements}, the instrument had to achieve a
spectral resolution $\delta\lambda=50-100$ m\AA.   
The choice of the LiNbO$_3$ etalon technology (developed by ACPO,
Australian Center for Precision Optics, CSIRO) was also clear from the beginning. 
By comparison, the more common
technology used in ground-based instruments of piezo-stabilized etalons (TESOS,
Kentischer \citeyear{Kentischer98}; GFPI, Puschmann \citeyear{Puschmann06}; and CRISP,
Scharmer \citeyear{Scharmer08}) poses serious demands in terms of weight of the
instrument and in-flight calibration, such as the need for calibrating  plate
parallelism and system finesses autonomously. Solid etalons based on man-made
LiNbO$_3$ crystals are comparatively speaking weightless and their calibration much
easier (see Section  \ref{sec:speccal}). The weight requirements of
this technology are driven by the pressurized enclosure (common to piezo
etalons) and by the high voltage needed by them to change the refractive index
of the crystal and tune between different wavelengths. A High Voltage
Power Supply (HVPS) working in the range of $\pm$5000 V has been employed (see
Section\ \ref{sec:PE}). Additionally, the relatively high refractive index of
this material ($n\approx 2.3$) permits incidence angles over the etalon $n$
times larger than their piezo-controlled counterparts, thus allowing for
smaller etalon sizes and more compact instruments.

LiNbO$_3$ etalons can be fabricated with thicknesses of 200 $\mu$m, or larger,
and with a reflectivity ($R$) high enough to permit resolving powers of 50
m\AA , as required, using just a single etalon. However, their fabrication finesse
($\textsf{F}_{\rm f}$, for a definition, see Atherton {\it et al.}, \citeyear{Atherton82}) is
hardly better than 30 over an aperture of six 
cm (Arkwright {\it et al.}, \citeyear{Arkwright05}).
As is well known (Atherton {\it et al.}, \citeyear{Atherton82}), it is convenient to use
reflective finesses $\textsf{F}_{\rm r}$ for the etalons similar to the
fabrication finesse, which combined provide a total finesse $\textsf{F}$ in the
range 20-25 ($\textsf{F}^{-2}=\textsf{F}_{\rm f}^{-2}+\textsf{F}_{\rm
r}^{-2}$).  The total finesse of the etalon, its resolving power, and the Free
Spectral Range (FSR) are related to each other by:
\begin{equation}
\textsf{F}={{\mathrm{FSR}}\over{\delta\lambda}}.
\end{equation}
Now, with $\textsf{F}=22$ (the value for the IMaX etalon) and
$\delta\lambda=50$ m\AA , the FSR is about one~\AA. Thus, to filter out
nearby etalon orders, a pre-filter with at least 0.5~\AA is needed
or, in its absence, a second etalon. None of these solutions is practical as
the smallest pre-filters available in the market have a FWHM of one \AA~and a
second etalon requires a second HVPS which would add considerable extra
complexity to the instrument. Instead, the solution adopted was to ask for an
etalon with $\delta\lambda=100$ m\AA~and have the light pass twice through it
(decreasing the FWHM by a factor $\sqrt{2}$). In this case, the FSR of the
etalon is raised to about two \AA~and a pre-filter with one \AA\ FWHM can conveniently
suppress the extra etalon orders.  In this way, a simplified instrument capable
of achieving the required resolving power, but still using
only one etalon, was defined. The exact properties of the pre-filter and etalon
are presented in detail below.

The exact location of an etalon in a magnetograph is to some extent a matter of
taste. The two options commonly used are to place the etalon near a telecentric
focal plane (TESOS, Kentischer, \citeyear{Kentischer98}; CRISP, Scharmer,
\citeyear{Scharmer08}) or at a pupil plane within the instrument (GFPI Puschmann,
\citeyear{Puschmann06}; IBIS, Viticchi\'e, \citeyear{Viticchie09}). The pros and cons of
both solutions have been discussed in some length by von der L\" uhe and
Kentischer (\citeyear{Vonderluhe00}), Gary {\it et al.} (\citeyear{Gary03}) and Scharmer
(\citeyear{Scharmer06}). Neither of the options is free of complex calibration
problems. IMaX selected the collimated configuration mainly because the exact
impact of the pupil apodization problem in magnetograms was not well understood
when the decision was taken.  The success of the CRISP instrument in achieving
diffraction limited snapshots (with the aid of blind deconvolution
post-processing) proves that this effect can be adequately controlled during
the design phase of a telecentric system.  In the case of the collimated
configuration, one benefits from a homogeneous instrument profile for all
points in the final image, but a wavelength shift across the FOV appears due to
the inclination of the off-axis rays as they reach the etalon. This effect is
commonly referred to as the wavelength blueshift of the collimated
configuration. Its calibration for the IMaX case is explained in Section
\ref{sec:obsanl}.

The biggest concern with an etalon in a collimated configuration is the
degradation of the image quality that it can produce.  In particular, von der
L\" uhe and Kentischer (\citeyear{Vonderluhe00}) derived an equation for the WFE
introduced by an etalon near a pupil plane due to the presence of cavity
(polishing) errors of a magnitude $\Delta_{\rm rms}$ (see their Equation 10). This
formulation predicted that it could be very difficult to reach
diffraction-limited performance in the collimated configuration (see also
Scharmer, \citeyear{Scharmer06}; Mart\'\i nez Pillet, \citeyear{Marpillet04}).  For a
LiNbO$_3$ etalon with $\Delta_{\rm rms}\approx1.3$ nm (corresponding to
$\textsf{F}_{\rm f}=30$, the value of the IMaX etalon over the full aperture)
and a typical coating reflectivity of 0.97, this equation predicts a wavefront
error of $\lambda/5$.  Such an optical element alone degrades considerably the
image quality of the system.  However, the WFE measurements made by ACPO/CSIRO and by
us using Zygo interferometers on real etalons (\'Alvarez-Herrero {\it et al.},
\citeyear{Alvarez06a}) have shown this equation to predict too large wavefront
degradations.  We have found two reasons for this large estimates.
First, Scharmer (\citeyear{Scharmer06}) pointed out that this equation has to be
averaged over the transmission profile and not only applied at the transmission
peak (where the wavefront deformation is maximum). Wavefront degradations will
be half as large when properly averaged over the transmission profile.
Second, and equally important, is the fact that when this equation is computed,
one should not use the value of the reflectivity in the etalon coatings.
Instead, one should infer an effective reflectivity that accounts for the total
finesse (reflective and fabrication) of the etalon and use this value in the
equation. That is, define $R_{\rm eff}$ so that
\begin{equation}
\textsf{F}={{\pi \sqrt{R_{\rm eff}}}\over{1-R_{\rm eff}}}
\end{equation}
and introduce $R_{\rm eff}$ instead of the coating reflectivity $R$ in the equation of 
von der L\" uhe and Kentischer (\citeyear{Vonderluhe00}). 
It is easy to understand why the effective reflectivity gives a better prediction
of the actual wavefront deformation. As shown by Atherton {\it et al.} (\citeyear{Atherton82}), the
total finesse is the number of effective bounces of the rays that get  
added coherently by the etalon to produce the interferometric pattern (instead
of the theoretical infinite number of them). Using an effective reflectivity, which  
correctly quantifies how many times the wavefront interacts with the etalon
surfaces, should produce a more accurate description of the deformation that it suffers.
Doing so, in the case of IMaX ($R=0.9$, $R_{\rm eff}=0.867$), the predicted wavefront error
is, at the transmission peak, $\lambda/14$ and when 
averaged over the full bandpass $\lambda/27$.
Measurements by ACPO/CSIRO gave values in the range $\lambda/23$-$\lambda/38$
for the two IMaX etalons (flight unit and spare). These values were confirmed
at INTA for the spare unit but not for the flight unit as
commented in section \ref{sec:iqpd}. In any case, it is clear that
diffraction-limited performance with an etalon near a pupil plane is 
indeed possible if
care is taken of the balance between cavity errors and coating reflectivity (for
the specific case of IMaX, see Mart\'\i nez Pillet {\it et al.} \citeyear{Marpillet04} and
\'Alvarez-Herrero {\it et al.} \citeyear{Alvarez06b}).

\subsubsection{Imaging}
\label{sec:imaging}

High spatial resolution was at the forefront in the list of requirements for
the definition of the mission and of the instrument. At system level, a
wavefront sensor with refocusing capabilities and low order aberration
correction was implemented (see Berkefeld {\it et al.}, \citeyear{Berkefeld10}). At
instrument level, and in order to cope with slow thermal drifts that change the
residual aberrations of the instrument, phase diversity (PD) capabilities were
included. IMaX uses a thick glass plate that can be moved in front of
the CCD camera at right angle to the beam entering the beamsplitter
(and referred to as camera 1; the camera in the same direction as the light
entering the beamsplitter is camera 2).  This plate is not inserted into the
beam in normal observing modes, but after an observing run is completed, the
plate is introduced and a burst of 30 images is acquired in the two cameras. These
images are observed only at continuum wavelengths and with no modulation
implemented in the LCVRs. However, the images are accumulated as in the
subsequent observing run to ensure a similar S/N ratio. Post-processing
analysis of these images allows the Zernike coefficients describing
the optical quality of the complete system to be recovered.

\subsection{Light and Data Path}
\label{sec:light}

Light reaches IMaX at the F4 focus of the ISLiD. The first element that is
encountered is the pre-filter (whose properties are given in Section
\ref{sec:pref}).  In this way only one \AA~ of solar flux enters the instrument,
while the rest is directly sent into a light trap. The pre-filter also protects
the LCVRs from any residual UV light that might still be present in the beam
(although the majority of the UV photons have been directed to the \Sunrise
UV Filter Imager, SUFI Gandorfer {\it et al.} \citeyear{Gandorfer10}).  
After the pre-filter, light travels through the two 
LCVRs (Section \ref{sec:lcvrs}) where the light intensity is encoded for the polarization
measurements. The pre-filter and the LCVRs share the same opto-mechanical,
thermally stabilized, mounting. Light then reaches the collimating lenses (the
optical design is described in Section  \ref{sec:optics}) and makes the first
pass through the LiNbO$_3$ pressurized enclosure (traversing, in this order, a
window, the etalon, and a second window). The etalon (Section  \ref{sec:etalon})
is inside a thermally stabilized enclosure provided by ACPO/CSIRO that included two
fused silica windows. Two 45$^\circ$ incidence folding mirrors turn the light
backwards to enter again the etalon enclosure and make the second pass through
it (in a different area of the etalon). An aperture stop (of 25 mm diameter) is
at a symmetric position between the two etalon passes. After exiting the etalon
enclosure for the second time, light passes through the camera lenses that
produce the final image on the focal plane detectors. The collimating and
camera lenses share the same opto-mechanical mounting located in front of the
etalon. The folding mirrors, at the other side of the etalon enclosure, use a
different mechanical mounting. The two housings on either side of the etalon
are thermally controlled with a proportional, integral, derivative system (PID;
the thermal design is described in Section  \ref{sec:thermal}). A third folding
mirror (located in the pre-filter and LCVR mounting) sends the light towards
the linear polarizing beamsplitter that finally directs the light to the two
CCDs (Section  \ref{sec:ccds}).  The motorized PD plate system lies in between the
beamsplitter and camera 1. The detectors read out are synced with the
LCVR states and the frames are sent to the real-time IMaX electronics (Section
\ref{sec:electronics}), where the images are accumulated $N_{\rm A}$ times. This is
repeated at each of the $N_\lambda$ wavelengths producing a total of $N_{\rm p}\times
N_\lambda$ images that are then sent to the IMaX on-board control computer.
Lossless JPEG-2000 compression is applied to the data that is, then, sent to
the \Sunrise ICU (Instrument Control Unit) and to the storage disks. Some
of the images can be sent as thumbnails to the ground control computers
depending on the available telemetry and the adopted thumbnail policy.
Polarization demodulation was implemented in the on-board software as
demodulated images have better compression factors. However this option was
finally not used during the flight as data storage capacity was never critical.

The LCVRs are tuned while the light is being recorded by the detectors. As
their tuning times are finite, the transitions of the LCVR states are
integrated by the detectors. The effect of this integration of non-constant
LCVR states is fully taken into account by the instrument polarization
calibration (described in Section  \ref{sec:polcal}). Wavelength tuning is slower
(see Section  \ref{sec:etalon}) and while the etalon changes the observed
wavelength, the CCDs are being read out, but the frames are discarded.

\section{Concept Implementation}
\label{sec:implementation}

\subsection{The IMaX Pre-filter}
\label{sec:pref}

   \begin{figure}
   \centering
   \includegraphics[width=9cm]{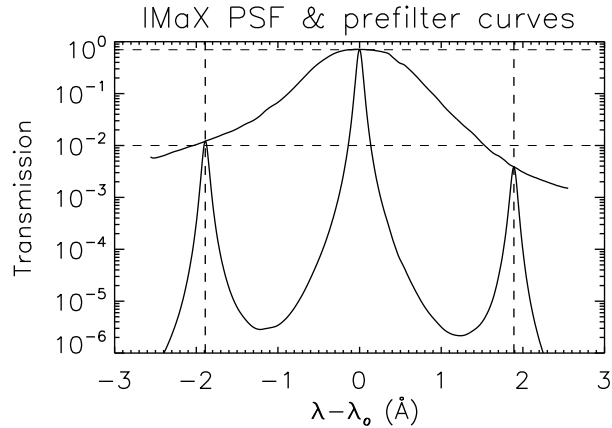}
      \caption{IMaX spectral point spread function. It includes an Airy
      function (three-peaked curve) for the etalon (in double pass) computed
      from empirical data provided by the manufacturer and multiplied by the
      Barr pre-filter bandpass curve measured at THEMIS (bell-shaped curve on
      top).  }
         \label{fig:imaxpref}
   \end{figure}

   \begin{table}
      \caption[]{IMaX Barr pre-filter characteristics}
         \label{tab:prefbarr}
         \vspace{-0.2cm}
     $$ 
         \begin{array}{p{0.5\linewidth}l}
            \hline
	    Property & {\rm Value} \\
            \hline
            \noalign{\smallskip}
		Central lambda  & 5251.15~\AA  \\
		FWHM  & 1.05~\AA \\
		Working $T$  & 35~^\circ {\rm C}  \\
		Peak transmission   & 0.65  \\
		Effective refractive index  & 1.92  \\
		Transmitted wavefront  & \lambda/4  \\
		Clear aperture  & 20~{\rm mm}  \\
		Temperature coefficient  & +0.025~\AA/^\circ {\rm C}  \\
		Angle for 5250.2  & 2.1^\circ  \\
		Design  & {\rm Double~cavity}  \\
		Construction  & {\rm Hard~coatings}  \\
	    \noalign{\smallskip}
            \hline
         \end{array}
     $$ 
\end{table} 

The availability of ultra-narrow bandpass, high transmission, filters from Barr
Associates was at the very core of the IMaX instrument definition. Metal-oxide
optical interference coatings are used in combination with various colored
glasses to produce filters with a bandpass of one \AA~and high out-band blocking
ratios. Additionally, these pre-filters have been demonstrated to be thermally
stable and environmentally robust.  The IMaX pre-filters (flight and spare
units) were taken to the French-Italian THEMIS telescope at Teide Observatory
for calibration. The bandpass was accurately measured at the nominal
temperature and is shown in Figure \ref{fig:imaxpref}. The properties of the
flight unit are given in Table \ref{tab:prefbarr}. These measurements
quantified three crucial aspects for the correct functioning of IMaX: the one
\AA~bandpass, a high transmission, and the sensitivity angle. As usual, the
pre-filter was ordered slightly shifted to the blue of the nominal working
wavelength; in particular, at 5250.65 +0.5-0.0 \AA~{\em air} wavelength. The
calibration performed in THEMIS showed that the angle needed for achieving
maximum transmission at 5250.2 \AA~was 2.1$^\circ$. A double cavity design was
selected in order to have a rapid fall off in the transmission wings of the
pre-filter so that the nearby etalon orders would be reduced below 1\%. While
this was achieved in the red wing side, the blue side of the pre-filter showed
an extended wing that left the secondary etalon peak there at a value slightly
above 1\%. The spectral calibration of IMaX (Section  \ref{sec:speccal}) has used
this pre-filter curve for its characterization.

During the flight, the heater power for the opto-mechanical mounting proved insufficient
and the nominal 35 $^\circ$C was actually never reached. The achieved temperature 
level was 30$\pm 1^\circ$C instead. Given the temperature coefficient of the 
pre-filter, this different temperature meant a bandpass drift to the blue of a
mere 0.1 \AA~ which has a negligible effect.

\subsection{The Liquid Crystal Variable Retarder Polarimeter}
\label{sec:lcvrs}

Two conceptually identical LCVRs were used as the polarization modulator in IMaX.
The LCVRs were produced by Visual Display, S.A. (Valladolid, Spain) in a
collaboration to produce space qualified LCVRs 
(in the context of the {\em Solar Orbiter} ESA
mission, see Heredero {\it et al.}, \citeyear{Heredero07}). 
IMaX's LCVRs (see Table \ref{tab:lcvrprop}) use voltages in the range of 0-10 V
to produce retardances between 535$^\circ$ (at 0 V) and of 20$^\circ$ at high
voltages. The voltages used for the vector modulation case were
$[2.54,2.54,3.12,3.12]$ V for the first LCVR and $[2.46,9.00,4.30,2.90]$
V for the second one. Each combination of voltages for the two LCVRs
produces each one of the modulation states, $I_i,~i=1,2,3,4$. The calibration of
voltages is explained in Section  \ref{sec:polcal}. Suffice it to say here that
they were selected to reproduce the canonical vector modulation case explained
in Mart\'\i nez Pillet {\it et al.} (\citeyear{Marpillet04}).  The longitudinal case used
$[2.292,2.292]$ V for the first LCVR, which corresponds to 360$^\circ$
retardance, and $[5.293,2.635]$ V for the second one, that provide the
$\pm\lambda/4$ retardances.  It is also important to note that long 
transitions from high voltages to small voltages are avoided as they are always
slower than transitions from low voltages to high ones. All transitions used
here showed response times below 40 ms. 

LCVRs are temperature sensitive. For IMaX, a nominal temperature of 35 $^\circ$C
controlled to within $\pm 0.5^\circ$C was specified. Since LCVRs share the mounting
and temperature control with the pre-filter, their nominal temperature was never 
reached and the whole system flew $5^\circ$C cooler. However, extensive 
laboratory calibrations performed with the IMaX LCVRs allowed us to know their
temperature sensitivity. The results from Heredero {\it et al.} (\citeyear{Heredero07}) have
been used to quantify the temperature sensitivity as:
\begin{equation}
{{\Delta\delta}\over{\Delta T}}=-1.16+0.305V-0.02V^2~~~~\mbox{for} ~V< 8,
\end{equation}
showing a tendency to produce smaller retardances at higher temperatures. $\delta$
is the LCVR retardance in degrees, $T$ is the temperature in degrees Celsius, and
$V$ the applied voltage in V. At voltages higher than eight V, the sensitivity to
temperature is negligible for these particular LCVRs.

   \begin{table}
      \caption[]{IMaX LCVRs characteristics}
         \label{tab:lcvrprop}
     $$ 
         \begin{array}{p{0.5\linewidth}l}
            \hline
	    Property & {\rm Value} \\
            \hline
            \noalign{\smallskip}
		Maximum retardance  & 780~{\rm nm}  \\
		Response time  & < 40~{\rm ms}  \\
		Working $T$  & 35~^\circ {\rm C}  \\
		Peak transmission   & 0.98  \\
		Transmitted wavefront  & \lambda/10  \\
		Clear aperture  & 25~{\rm mm}  \\
		Temperature coefficient  & -1.16~{\rm deg}/^\circ {\rm C}  \\
	    \noalign{\smallskip}
            \hline
         \end{array}
     $$ 
\end{table} 
   \begin{figure}
   \centering
   \includegraphics[width=9cm]{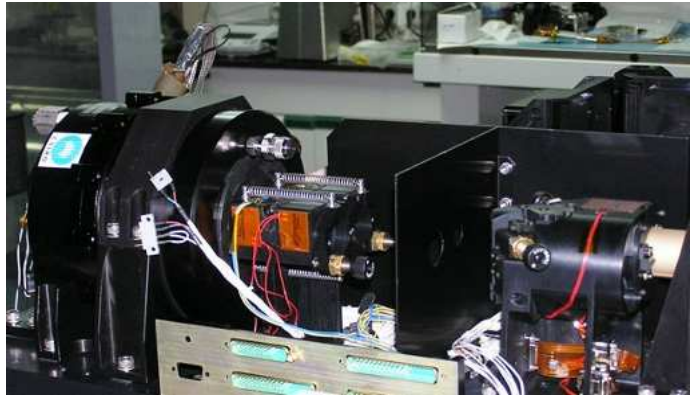}
      \caption{The temperature controlled side of the IMaX optical bench
      including the pre-filter and LCVR mounting (right), the collimator and
      camera lenses mounting, and the etalon TCPE (left). The stray-light baffle
      is seen in between the two first elements.  }
         \label{fig:imaxhot}
   \end{figure}

\subsection{LiNbO$_3$ Etalon}
\label{sec:etalon}

Solid LiNbO$_3$ etalons have many advantages for their use in autonomous
platforms (low weight and parallelism stability among others). However, the most important
disadvantage for their implementation in a balloon platform is the use of high
voltages for their tuning with the consequent possibility of producing an
electrical discharge in the system.  The flight pressure conditions were near 3
mbar, exactly the region were air ionization occurs more easily (for both space
or ground conditions this is a less severe problem). We thus identify the
possibility of an electrical discharge in the kV LiNbO$_3$ system as one of the
design drivers for the instrument. The implications were that both the etalon
and the HVPS had to be in pressurized chambers located in the optical bench.
The connectors between the HVPS and the etalon also deserved special care and
avoided the use of outer metallic components. 

The LiNbO$_3$ etalons were ordered early in the project because of their
relatively long manufacturing time ($\approx$ one year). ACPO/CSIRO built two etalons
inside thermally controlled pressurized enclosures (TCPE) containing a dry
nitrogen atmosphere inside, at 1.2 bar. The TCPE internal atmosphere was
controlled to within $\pm 0.03~^\circ$C and was specified to work at 35
$^\circ$C. This susbsystem power was sufficiently dimensioned to reach this
temperature with enough margins all along the flight time. The TCPE was ordered
with high-quality-grade fused silica windows of 14 mm thickness and 70 mm
diameter. The predicted deformation of the windows once the TCPE was in near
vacuum conditions was of 380 nm peak to valley over the full aperture. The
windows were anti-reflection coated to have almost perfect transmission at 525
and 633 nm. 

The two etalons acquired to ACPO/CSIRO are referred to as TCE-116 and TCE-117. 
The best of them in terms of surface polishing is the TCE-116 etalon, but its TCPE 
developed some leakages through the high voltage connectors and could not be
repaired in time for the flight. The TCE-117 was finally installed and the one that
flew in IMaX. Its properties are given in Table \ref{tab:acpoetalon} (the most 
substantial difference between TCE-117 and TCE-116 is that the latter had a full 
aperture wavefront error in single pass of only $\lambda/38$). The properties given in 
that table refer, unless otherwise stated, to any sub-aperture of 25 mm instead of the 
full aperture. IMaX pupil image near the two etalon passes is of this size, or 
slightly smaller, so the requirements to ACPO/CSIRO were given over any 25 mm subaperture. 
The combination of the IMaX pre-filter and the etalon transmission curve in double
pass (the instrument spectral profile) is shown in Figure \ref{fig:imaxpref}. 
The pre-filter wings decrease the secondary maximum from the side
orders of the etalon to levels of around 1-2\%. The final spectral resolution
achieved by the instrument and the spectral calibration are explained in Section 
\ref{sec:speccal}.

The TCPE cages include a tilt motor to change the incidence angle of the
etalon. While this motor was very useful during the AIV phase, the etalon
was run at near normal incidence during the whole flight.

   \begin{table}
      \caption[]{IMaX LiNbO$_3$ etalon properties (25 mm aperture)}
         \label{tab:acpoetalon}
     $$ 
         \begin{array}{p{0.5\linewidth}l}
            \hline
	    Property & {\rm Value} \\
            \hline
            \noalign{\smallskip}
		Construction  & z-{\rm cut}  \\
		Aperture  & 60~{\rm mm}  \\
		Refractive index & 2.3268\\
	    	Thickness & 281 ~\mu{\rm m} \\
		$\textsf{F}_r$ ($R$)  & 27.5~(0.892)  \\
		$\textsf{F}_f$ ($\Delta_{rms}$)  & 35.8~(1~{\rm nm})  \\
		$\textsf{F}$ & 21 \\
		$\delta\lambda$ (single pass) & 93~{\rm m\AA} \\ 
		Tunning constant &  0.335\pm0.002~{\rm m\AA/V}\\
		Tunning speed  & 1500~{\rm V/s}  \\
		Working $T$  & 35~{\rm ^\circ C}  \\
		Temperature sensitivy   & 25.2~{\rm m\AA/{^\circ}C}   \\
		Temperature stability   & 0.03~{\rm ^\circ C}   \\
		Transmitted wavefront (single pass \& windows, full aperture) & \lambda/23  \\
		Homogeneity (rms fluctuations of peak $\lambda$)& 16~{\rm m\AA}  \\
	    \noalign{\smallskip}
            \hline
         \end{array}
     $$ 
\end{table} 

Perhaps the largest inconvenience of the LiNbO$_3$ technology for solar
observations is the tuning speed.  ACPO/CSIRO recommends to use tuning speeds smaller
than 1500 V s$^{-1}$ or 0.5 \AA~s$^{-1}$ (because of the piezo-electric effect of this
crystal could damage the wafer at higher tuning rates). As the typical steps
used in IMaX for spectral tuning are of about 40 m\AA, the etalon spends 80 ms
tuning the wavelength. For jumps of 0.4 \AA, that is, the jump corresponding to
switching between the continuum point and the first point inside the line, it
takes almost a second. This time is relatively long if we compare it with
piezo-stabilized systems that use  10 ms or so for virtually all tuning steps.
We tested a raw LiNbO$_3$, not polished to etalon standards, for tuning speeds
of 3000 V/s during 15 days satisfactorily (no wafer breakdown or degradation).
However, and in order to reduce risks, the tuning speeds during this first flight
were kept at the recommended rate of 1500 V/s. 

\subsection{Focal Plane Detectors}
\label{sec:ccds}

While the original plan was to fly two custom-made high quantum efficiency
detectors, these plans were abandoned as the requirements for a vacuum capable
detector similar to those that are state-of-the-art in ground instruments were
too stringent.
Instead, a thermally robust, vacuum compatible, front-side illuminated detector
was finally selected for IMaX. Two DALSA Pantera 1M30 cameras with 1K$\times$1K
chips (see Table \ref{tab:ccdprop}) were finally integrated. A frame transfer
design with a transfer time of four ms allowed us to avoid the use of any shutter
mechanism that would have complicated the instrument.  The chip itself is
protected by a one mm thick window in front of it.  The camera was mainly
selected for its robustness from a thermal point of view. The chip runs hot (at
45 $^\circ$C) and uses no active cooling mechanism.  Thermal control is based
on conduction to the front surface of the housing were it gets dissipated.  The
cameras have no special enclosure for vacuum conditions and have consequently
been extensively tested before the flight.  One DALSA camera was put in a
vacuum cryostat for 18 days working at four frames per second (fps).  Thermal
conduction was performed using an interfiller and a heat stripe connected to
the cryostat housing. Cooling of the cryostat was done by a Peltier mechanism
(replaced by a radiator during the flight). The camera survived without any
performance degradation and the decision of using them in IMaX was taken. The
small quantum efficiency ($Q=25$\%) could be considered acceptable only thanks
to the confirmation we had by then (after preliminary design review) of the
high transmission of the Barr pre-filter (see Table \ref{tab:prefbarr}).  It is
however clear that a higher QE camera would have allowed us to obtain more points
within the spectral line in the same interval of time (something that would have
also required the corresponding communication buses to cope with the 
higher frame rates).

\begin{table}
      \caption[]{IMaX focal plane detectors}
         \label{tab:ccdprop}
     $$ 
         \begin{array}{p{0.5\linewidth}l}
            \hline
	    Property & {\rm Value} \\
            \hline
            \noalign{\smallskip}
	   Chip & {\rm FTT1010M}   \\
	   Format & {\rm Frame~transfer} \\
	   Max. frame rate & 30~{\rm fps} \\
	   Data format & 12~{\rm bit}~(4092)\\
	   Pixels & 1024\times 1024  \\
	   Pixel size	 & 12~\mu{\rm m}  \\
	   Full well 	 & 170000~{\rm e}^-  \\
	   QE & 25 \% \\
	   Readout noise & 50~{\rm e}^-  \\
	   Dark current & 400~{\rm e}^-~{\rm s}^{-1}{\rm /pixel}  \\
	   Gain & 50~{\rm e}^-~{\rm DN}^{-1}\\
	   DC offset & 50~{\rm DN}\\
	   Frame transfer time & 4~{\rm ms}\\
	    \noalign{\smallskip}
            \hline
         \end{array}
     $$ 
\end{table}

\section{Optical and Opto-mechanical Design}
\label{sec:optics}
   \begin{figure*}
   \centering
   \includegraphics[width=9cm]{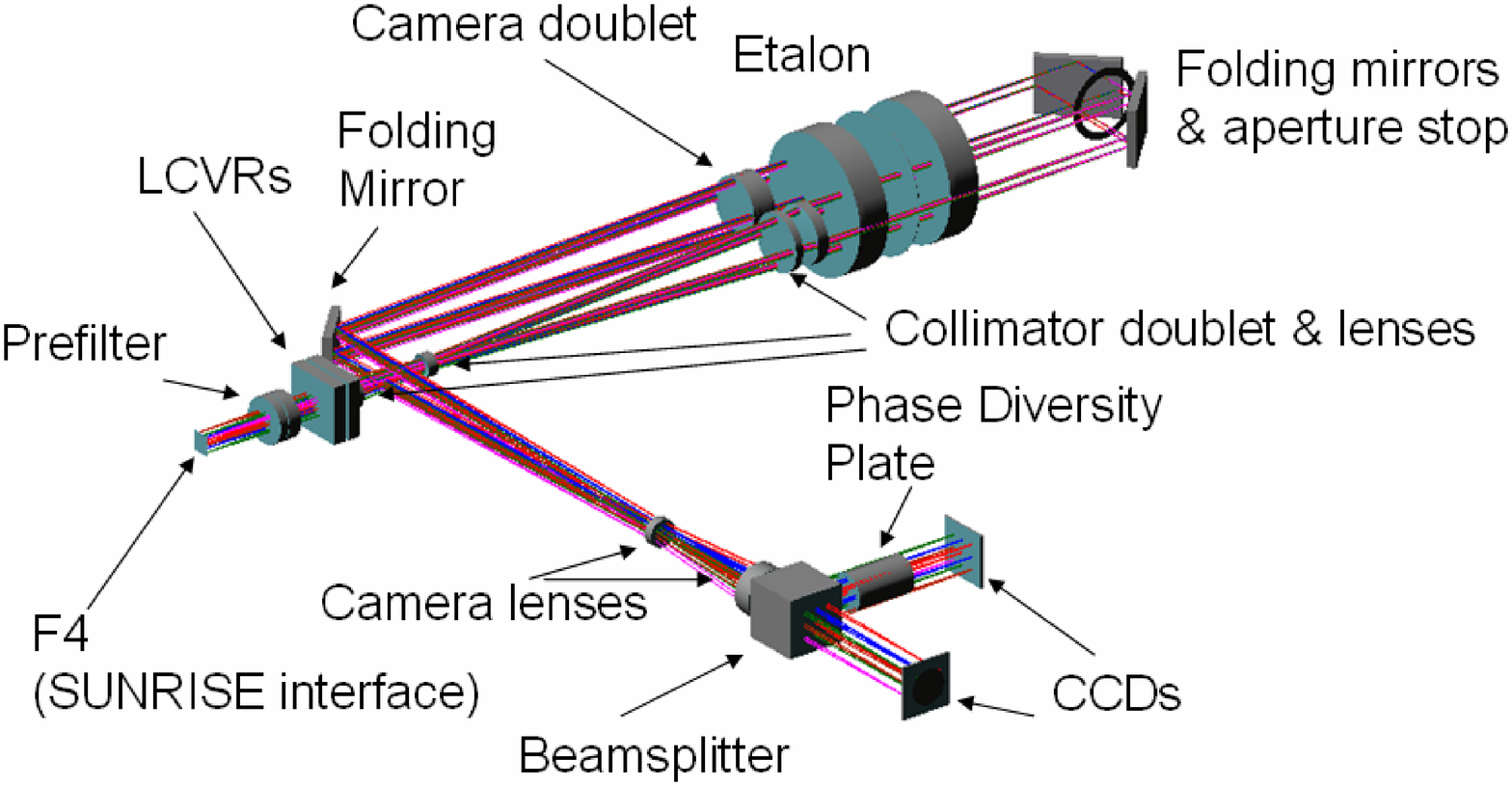}
   \includegraphics[width=9cm]{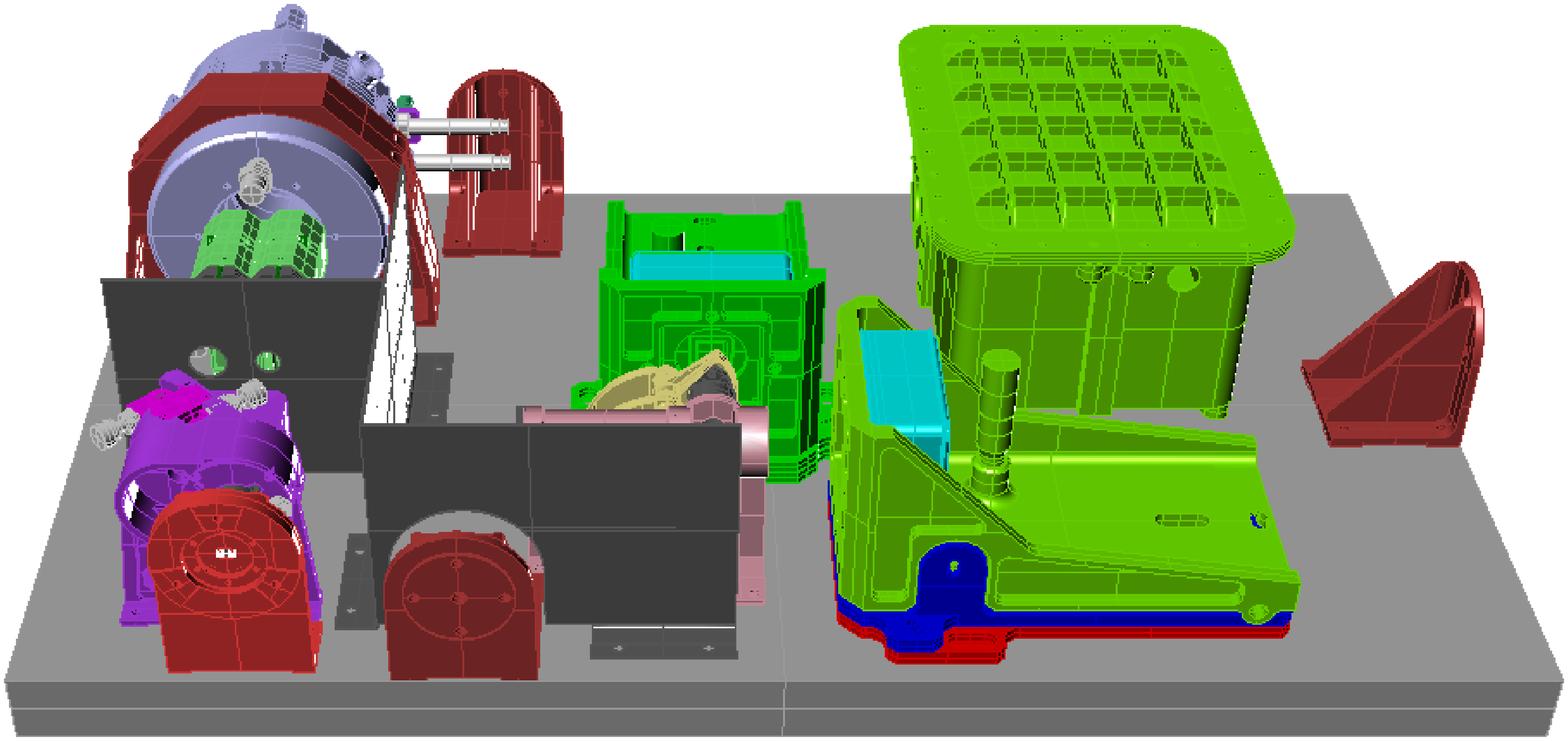}
      \caption{IMaX optical (top; self-explained) and opto-mechanical (bottom)
      design. In the latter, the F4 plane can be distinguished in brown-reddish
      color at the lower left corner. Just after it, the pre-filter and LCVRs
      housing can be seen in purple. After the stray-light baffle (dark grey),
      the etalon housing is in blue, preceded  by the collimator and camera
      lenses housing in green. The double light path can be discerned through
      the two holes in the baffle. To the right-hand side of the etalon
      enclosure, the vacuum HVPS connectors can be seen in white. After the
      central baffle, the beam splitter housing (in pink) appears. Behind it,
      in yellow is the PD mechanism. In green are the mountings for the two
      cameras (blueish) and the PEE. Three isostatic mountings (in brown) for
      integration into the instrument platform complete the design.}
      \label{fig:optics3}
   \end{figure*}

The IMaX optical power system is all refractive, the mirrors of the system
being used only for folding and packaging (see Figure  \ref{fig:optics3}). Three
mirrors in total are needed.  The optical interface with \Sunrise is the
ISLiD focal plane F4 (see Gandorfer {\it et al.}, \citeyear{Gandorfer10}).  At F4, a field
stop confines the IMaX FOV and prevents unwanted light from entering the
instrument and generating parasitic light. This field stop admits two focus
positions corresponding to the air and vacuum focus positions that differ by
1.64 mm in IMaX (the latter being displaced towards the instrument).  This shift
between the air and vacuum conditions was both predicted by the optical design
and checked during the vacuum tests (see Section  \ref{sec:vacuum}). Next,
the pre-filter and LCVRs mounting follow at a distance from F4 of about 30 mm
(where the focal depth is $\pm$0.66 mm; near the CCDs it becomes $\pm$2.1 mm).
After polarization modulation the beam goes through a collimator system
consisting of a doublet and two lenses.  The focal length of this collimator is
567.36 mm and its F-number is 25, matching the \Sunrise Telescope and ISLiD
F-number. The etalon works in a collimated space producing a blueshift over the FOV.
The collimated set-up for the etalon requires special care with the maximum
angle of incidence on it.  The focal length of the collimator guarantees that
the angle of incidence on the etalon does not exceed 0.44$^\circ$, which produces a
wavelength drift of 28 m\AA\ (over the largest circular FOV).  The diameter of
the resulting collimated beam is 25.6 mm, representing the actual area used of 
the etalon in each pass.  In the collimated space between the two folding
mirrors, we locate an aperture stop to adequately set the entrance and exit pupils of
the system.  The beam is finally focused onto the CCDs by a camera consisting
of a doublet and two lenses, with a camera focal length of 1021 mm
producing a final image IMaX F-number of 44.99.  The image focal plane of IMaX has
been designed to be telecentric and so has the object focal plane at F4. Both
sub-systems, the collimator and the camera optics are telephoto lenses in order to
shorten the total length of the system and get reduction ratios of 0.53 and
0.57, respectively.  The magnification of IMaX is 1.799 in order to get a final
image scale of 0.055 arcsec pixel$^{-1}$.  The polarizing beam splitter is
commercial and made of NBK7-Schott with anti-reflection coatings on all
surfaces.  In order to perform phase diversity correction on the image, a
parallel plate can be inserted in one of the channels in order to defocus the
image on this channel (camera 1).  
This plate is made of fused silica with 27 mm thickness that produces the specified 
amount of defocusing corresponding to a phase shift of one $\lambda$ 
at the edge of the pupil. There are no mechanisms inside the
instrument for maintaining IMaX in focus during the flight. For this reason,
the most common glass used in IMaX is fused silica which has near zero thermal
expansion (and low stress birefringence). To test and verify the optics of IMaX
with a Zygo interferometer (at 632.8nm), the system has been designed for a
wavelength range that includes that of the He-Ne laser. IMaX has thus been
achromatized for the range 524.9nm to 632.8nm.  The materials choice has a
great impact on both the achromatization and the athermalization of the
instrument. So, for the IMaX doublets a pair of glass materials that
compensate, both, the chromatic aberration and the sensitivity to changes in
temperature was selected.

The influence on the PSF of retroreflection on optical surfaces (ghost images)
was analyzed in detail. IMaX includes a high variety of optical components and
properties which calls for performing a thorough study on the influence of each
surface on the final performances of the instrument. Some of the optical
elements, like the etalon and the pre-filter, were considered in a simplified
way to reduce the time needed for the stray-light simulation, but maintaining
the strength of the signal. Since the coherence length of light going out from
the etalon is quite significant (54 mm for a 50 m\AA\ spectral resolution),
ghost beams should be added in amplitude and not only in intensity.  The flux
obtained in the detector integrated for the total PSF (including ghost images)
has been compared with that obtained from the ghost image alone.  The main
conclusion of our ghost-image studies was the necessity to tilt the etalon TCPE
by an angle of 0.36$^\circ$ around the horizontal axis perpendicular to the light
path in order to get rid of the ghost images produced by its windows. 

The IMaX opto-mechanical system consists of an optical bench where all the
mechanical and optical components are distributed
(see Figure \ref{fig:optics3}). To facilitate
mechanical manufacturing, all the sub-systems are independent modules. Thus,
the instrument contains a mounting for each of these sub-systems:
\begin{enumerate}
\item F4 diaphragm
\item Prefilter and LCVRs
\item Collimation and camera lenses
\item Etalon TCPE
\item Folding mirrors
\item Beam splitter
\item PD mechanism
\item CCD cameras
\item Stray-light baffle
\end{enumerate}

The F4 diaphragm has two possible orientations at 180$^\circ$ from each other.
In one of the orientations, the field diaphragm is in the air focus position as
seen from the CCDs whereas the other orientation corresponds to the vacuum
(flight) focus position.  The drawback in the individual mountings policy is
that one needs to align each of these mountings carefully. The optical axis was
defined as the normal to F4 passing through its centre. The integration process
consisted of the individual assembly and verification of each module by
sequential order, from F4 to the CCD cameras. This was done with the help of a
control metrology machine assuring a mechanical positioning accuracy better
than 25 $\mu$m and higher alignment accuracy was achieved with the help of a
theodolite (up to two arcsec).  Once all the mechanical elements and mirrors were
assembled, the power optics (lenses) were mounted.

A stray-light baffle was included in the instrument concept
in order to reduce the view factor of the cameras with respect to the zones 
with maximum scattering power. The baffle can be seen in Figs \ref{fig:imaxhot} and
\ref{fig:optics3}. All the design calculations have been made with the ASAP code.

\section{Instrument Electronics} 
\label{sec:electronics}

\begin{figure}
\centering
\resizebox{0.7\hsize}{!}{\includegraphics[width=\textwidth]{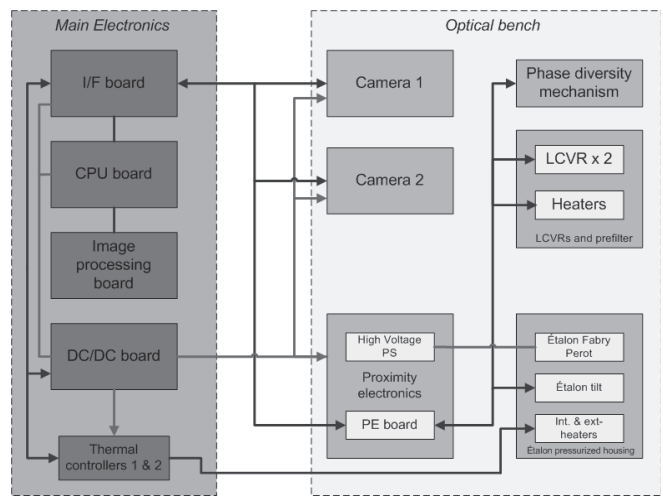}}
\caption{IMaX electronics block diagram.} 
\label{fig:E-BD} 
\end{figure}

The IMaX electronics are made up of three main blocks, namely, the main
electronics (ME), the optical bench electronics, and the harness. In the
optical bench, a proximity electronics (PE), the CCD cameras, and the
mechanical, thermal, and optical actuators are located. Everything is
manufactured with commercial-grade, either off-the-shelf (COTS) or specifically
designed, components. Both the ME and the PE are enclosed in pressurized
vessels (MEE and PEE, respectively) that provide ground pressure conditions 
during the flight. The harness
envelopes are chosen to minimize the outgassing impact in the quasi-vacuum
flight conditions. 
The present section is devoted to explaining the various
functionalities and the overall design considerations of these three blocks. We
do it separately in subsections since they are physically and functionally
distinct. A block diagram of the IMaX electronics can be seen in Figure\
\ref{fig:E-BD}.

\subsection{Main Electronics}
\label{sec:ME}

The ME (see bottom panel of Figure \ref{fig:IMaX}) are responsible of the overall control 
and data processing of the
instrument and of the interface communications with \Sunrise.
Their enclosure is physically located on a rack
along with the other instrument electronic boxes and the platform control
electronics.

The main functionalities of the IMaX ME are ensured through five sub-systems,
namely, the CPU board, the real-time image processing board (IPB), the interface (I/F)
board, the etalon thermal control, and the power supply (DC/DC board). 

The CPU board is a COTS VL-EBX-12b board from Versalogic Corporation,
that uses a PC/104 expansion bus and a SDCFX3-4096 solid state, massive
storage, four GB disk from SanDisk that includes the operating system and all the
specifically designed software (see Section\ \ref{sec:SW}). It has the following
features:
\begin{itemize}
\item Pentium 3 Mobile microprocessor @ 1.6 GHz.
\item 512 MB RAM.
\item Four GB, compact flash memory disk.
\item Two Ethernet 10/100 ports.
\item One PC/104-Plus slot used by the SMT374-300 board.
\item Two COM ports (RS-232) + two COM ports (RS-422).
\item Eight-channel, analog-to-digital converter with 12-bit resolution.
\item 32 channel digital input/output.
\item Graphic video.
\item Watch-dog timer.
\end{itemize}
This board is in charge of
\begin{itemize}
\item booting the system from a NVRAM memory and from the disk,
\item initializing the different sub-systems,
\item controlling the different sub-systems,
\item carrying out watch-dog tasks,
\item communicating with the PE and the IMaX optical bench,
\item switching on and off the secondary power supplies,
\item compressing images to be stored,
\item and communicating with the \Sunrise Instrument Control Unit (ICU).
\end{itemize}

The IPB is implemented through a mezzanine board with a SMT-310Q PC104
carrier and a Sundance SMT374-300 board. The latter includes two Texas
Instruments C6713 DSPs (digital signal processors) and one Virtex-2 XC2V2000
FPGA (field programmable gate array) from Xilinx. The board runs at 200 MHz and
has two SDRAM banks of 128 MB each. DSP{\#}1 is in charge of the communications
with the embedded CPU and the PE. It also has demodulation and truncation
capabilities of the accumulated images. DSP{\#}2 performs 
image accumulation according to the different observing modes and of sub-systems
(etalon, LCVRs, CCDs, phase diversity mechanism) synchronization. The
communications between this board and the CPU, the PE, and the CCDs are carried
out through the FPGA. 

The interface board has been introduced for mainly historical reasons.
As commented on earlier, the first customized cameras chosen for IMaX 
had to be changed in a quite
advanced step of the design due to a failure of the supplier
in providing them such that they complied with the specified
requirements. To keep the on-line processing developments as unchanged as
possible, the easiest solution at that point was to introduce this board that
simulate the foreseen behavior of the former cameras with the COTS ones.
Therefore, all the communications with the cameras are indeed made through this
board that also controls the housekeeping signals conditioning of the main
electronics box.

The functionalities of the remaining two sub-systems are fairly clear. The
etalon thermal controller maintains the interferometer within
the specified range of temperatures. It is implemented with one PID
2216e controller from Eurotherm that provides an
accuracy of $\pm 0.03^\circ$C. A second PID device controls the ambient surrounding the
TCPE to within $\pm 0.5^\circ$C. The specifically designed power
supply feeds the different sub-systems and the MEE fan. The supply is made from a  
DC non-regulated voltage provided by the
platform, taking care of the necessary efficiencies to prevent power losses,
allowing soft starts for in-rush current minimization and sequential switch-on
of the sub-systems, and with the necessary filters to maintain electromagnetic
compatibility with the other instruments aboard \Sunrise. All these
features follow a strict grounding and bonding scheme (similar to that in space
applications), hence avoiding current returns.

\begin{figure}
\centering
\resizebox{0.7\hsize}{!}{\includegraphics[width=\textwidth]{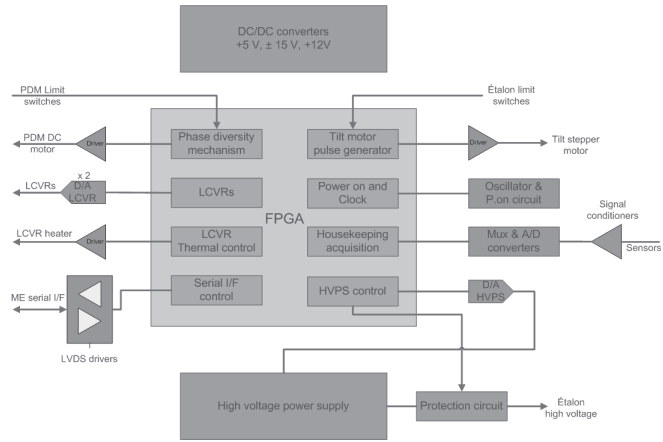}}
\caption{IMaX proximity electronics block diagram.} 
\label{fig:PE-BD} 
\end{figure}

\subsection{Optical Bench Electronics}
\label{sec:PE}

Due to the long cabling distances between the MEE and the OBE, proximity
electronics have been placed in a pressurized enclosure near the optical
components mainly because of the presence of the HVPS. We took advantage of the
PEE to install additional electronics in order to avoid voltage drops, noise
generation and reduce the number of wires, reading errors, {\it etc.} The main
functions of the PE are:
\begin{itemize}
\item the LCVR signal generation and thermal control, 
\item the etalon tilt motor and high voltage power supply control, 
\item the synchronization between changes of the etalon wavelength sample and the LCVR retardation angle, 
\item the acquisition of the optical bench housekeeping parameters, 
\item the power drive for the heaters, 
\item the generation of the different voltages for the sub-systems, 
\item the reset and clock generation, 
\item and the communications with the ME. 
\end{itemize}
All these functions are carried out by a specifically designed board using a
non-volatile 1280 AX FPGA from ACTEL as a main controller for peripheral
devices such as the HVPS or the controllers for the LCVRs and the phase
diversity mechanism. 

A block diagram of the PE is shown in Figure\ \ref{fig:PE-BD}. The DC/DC
converters are included in order to avoid several different voltages traveling
between the ME and the PE through a long harness. High efficiency DC/DC
converters have been selected, hence minimizing power losses. The low-voltage
differential signaling (LVDS) driver block is formed by transmitter/receiver pairs
compatible with the LVDS electrical standard. The communications with the ME are 
carried out by the FPGA through customized, serial-synchronous interface with
five lines from the ME and two lines to the ME. The block labeled MUX and A/D includes
an analog multiplexer and a 12-bit resolution analogue-to-digital (A/D) converter with a range
between 0 V to + 10 V. It constitutes a 16-channel acquisition system for
temperatures, pressures, and voltages. The signal conditioners are in charge of
adapting the different analogue voltages to the range of the A/D converter. The
LCVRs D/A converter module has two devices to generate the analogue signals
needed to control the retardance in both liquid crystals. A one kHz, square
signal voltage modulated is generated between the FPGA and this module with an
amplitude of $\pm 15$ V. Applied to the LCVR, this signal produces a retardance
controlled by the amplitude of the applied voltage.
This signal should later be adapted in impedance by an
LCVR analogue driver. A similar D/A converter is in charge of controlling the
HVPS. A COTS, HP5RZC bipolar power supply between $\pm
5000$ V from Applied Kilovolts, Ltd. has been used. The output of this HVPS is
linear to the input. After calibration, $V_{{\rm out}} = 448.22 \, V_{\rm in}
+ 13.14$ with all quantities in V. The input range goes from $-10$ V to $+10$ V. 
The converter
resolution is 16 bit, thus providing the required accuracy for voltage control.
To prevent any damage in case of power supply failure, a circuit using a high
voltage relay has been included that allows a slow discharge of the etalon
when the supply is off.

The cameras (see Section\ \ref{sec:ccds})
are able to reach up to 30 fps with a 12-bit resolution. The nominal speed
during flight was four fps. A serial interface based on channel link is used
allowing to employ just four fast lines for transmission and one line for
triggering plus two slow lines for serial control. Two operating modes (out of
the possible six) are used, namely the external frame rate and exposure
time mode for integration times greater than one second (only used for assembly,
integration, and verification procedures), and the external frame rate and
internal exposure time mode for integration times less than one second. Synchronism
between the two cameras is allowed within both operating modes. Commands are
sent to the cameras by following a serial, asynchronous interface standard with
hardware handshake.

Two motors are installed inside IMaX but they are not used under normal
observational conditions. They are only used for calibrations, either during flight or for the various ground-based check-outs. 
The phase diversity mechanism is actuated by a Maxon DC motor operated at 12 V.
Two position limit sensors report on the state of the bi-stable device. This motor
was extensively used during the flight. The
proprietary linear actuator, inside the TCPE, controlling the tilt angle of the 
etalon is based on a stepper motor driven by the PE and it was provided by ACPO/CSIRO.
This capability has been used during the assembly,
integration, and verification phase of the development but not during the flight when
the etalon was always at normal incidence.

\subsubsection{Harness and Connectors}
\label{sec:harness}

Data and signal links between the electronic boxes use differential techniques
with twisted pair wires for best common mode noise rejection and ground loop
prevention. Harness are protected against electro-static discharge effects and
radiated electromagnetic emission and susceptibility by overall shields. Theses
shields provide full coverage of all internal wires and inner shields. The
coverage includes stress relief areas at the connector backshells. The overall
shields are electrically connected to the structure ground level.  To provide
well defined low impedance ground connections, overall shields are attached to
connector pins on both harness ends. In that case the dedicated connector pins
of the unit or box receptacle are connected to the structure ground via short
($< 5$ cm) straps, in order to avoid loop-antenna-induced high
frequency noise. Harness and connectors outside the pressurized boxes are
vacuum compatible. Harness insulators are made of polytetraflouroethylene
derivates. Polyvinil chloride is avoided for outgassing reasons. Connector crimp contacts are
used. Soldered contacts are avoided because of the difficult cleanability of
flux contaminated wires. Cables, especially those inside optics compartments
and on the telescope are cleaned and baked out before processing or harness
manufacturing. All external connectors to the ME are space-qualified D-sub
connectors plugged to vacuum feed-throughs installed in the enclosure walls.
Vacuum circular connectors from LEMO have been used for the PE.

\subsection{Instrument Software}
\label{sec:SW}

The IMaX flight software is divided in two main blocks, the control software
(CSW) and the image processing (or DSP) software (DSW). The CSW is mainly
devoted to image compressing and formatting, telecommand and telemetry
management, and sub-system control. The DSW mainly deals with image
acquisition, (on-demand) demodulation and truncation, hardware (HW) control, and
interfaces with the CSW. Thorough validation procedures, similar to those required
by the European Space Agency for space missions, were applied to the system
after every change in any piece of SW during the assembly, integration, and
verification phase.

\subsubsection{Control Software}
\label{sec:csw}

\begin{figure}
\centering
\resizebox{0.6\hsize}{!}{\includegraphics[width=\textwidth]{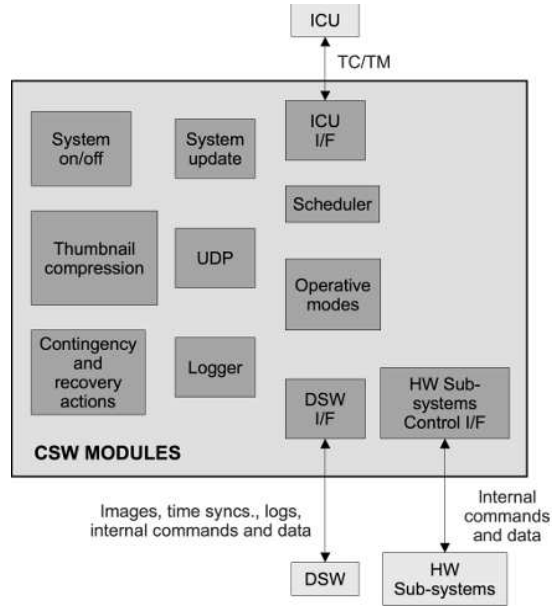}}
\caption{IMaX control software functional diagram.} 
\label{fig:CSW-BD} 
\end{figure}

A functional diagram of the CSW is shown in Figure\ \ref{fig:CSW-BD}. The system
on/off module controls the start up and shut down processes of IMaX. The system
update module implements software upload capabilities if required 
during the flight. The module labeled UDP manages the
acceptance, validation, and execution of user-defined commands and programs (see
below). The scheduler handles the (periodic or single) event tasks like housekeeping parameter (HK) 
generation and contingency detection. A specific module is devoted to detecting
contingencies and running the appropriate recovery actions. The logger module
is in charge of writing a set of files with all the relevant system actions and
events. Another module changes the various operative modes (see below).
Finally, three I/F boxes can be discerned in the diagram corresponding to ICU,
hardware, and DSW interfaces. Telecommands (TC) and telemetries (TM) are sent
and received through the first of these interfaces; internal commands and data
travel through the other two; besides, time synchronizations and log parameters
stream through the DSW I/F. Windows XP has been finally selected as the most
suitable operating system for all the IMaX purposes mostly because the
requirements of SW development for the SMT374-300 board. All the non-critical
services of the operating system are disabled and those files that are not used
are removed in order to minimize the start up time.

Three main concepts are involved in the IMaX SW development, namely, the
context file (CF), the so-called defined commands (DECOs) and user-defined
program (UDPs), and the operative modes. A CF is an archive containing all the
input system variables like the default observing mode and its consequent
values for wavelength samples and LCVR voltages, maximum and minimum allowable
values for, {\it e.g.}, temperatures in order to trigger contingency actions, {\it etc.}
There is always a default CF stored in the CPU board storage disk which is used
provided no other CF has been sent from the ICU (automatic time-line operation)
or the ground support equipment (remote operation). All the variables in the
operating CF can be modified either individually or in unison through
appropriate TCs. A set of specifically designed commands (DECOs) constitute the
basics of the CSW and represent all the necessary elementary actions needed to
operate the instrument and debug the CSW itself. They can be grouped as
follows:
\begin{itemize}
\item General DECOs: they are used for the CSW syntax ({\it e.g.} loop creation), for calling UDPs (see below), write comments, {\it etc.}
\item System DECOs: they are aimed at controlling the system ({\it e.g.}, start up, shut down, reset, {\it etc.})
\item Sub-system DECOs: they control the various functionalities of the sub-systems like the cameras, the PE, and the thermal controllers.
\item Operating and observing mode DECOs: they provide the configuration values for each mode.
\item HK DECOs: they ask for the several HK parameters. 
\item CF DECOs: they are in charge of modifying or verifying the CF.
\item Debugging DECOs: they are needed to build meaningful UDPs.
\end{itemize}
An UDP is just a collection of DECOs and represent a given high-level
action like constructing a piece of the observation time-line. There are three
operative modes for the CSW, namely, the safe, change, and science modes. The
system is in safe mode either when just after start up, just before shut down,
or upon request by the user. In this mode, the OB electronics are off and UDPs
can be run. In science mode the instrument performs all the automatic
activities corresponding to the current scientific observing mode. No UDPs can
be run in this mode. Any change between these two operative modes is preceded
by putting the system in the so-called change mode. When in this mode, manual
operations like those during the assembly, integration, and verification phase
can be carried out because UDP running is allowed.

\subsubsection{Image Processing Software}
\label{sec:DSW}

\begin{figure}
\centering
\resizebox{0.4\hsize}{!}{\includegraphics[width=\textwidth]{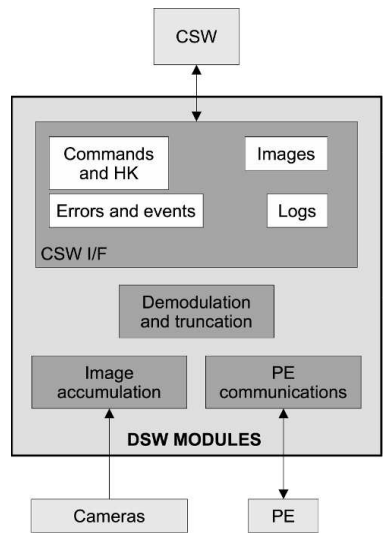}}
\caption{IMaX DSP software block diagram.} 
\label{fig:DSW-BD} 
\end{figure}

The main functions of the DSW are sketched in Figure\ \ref{fig:DSW-BD}. Among the
communications with the CSW one can find the reception of and response to
general commands and those concerning the HK; the sending of pre-processed
images; information on errors and other special events; and the sending of
logs. The DSW is also in charge of image acquisition and accumulation in the
different observing modes. These processes are carried out in real time, so
that the accumulation and exposure times are equal. Periodic triggering of the
cameras is made with an accuracy of $\pm 1$ ms and with a period equal to the
exposure time plus four ms (camera internal transfer plus erase times). All the
accumulations are made with 16-bit words.
The communications with the PE consist of the transfer of values
corresponding to the etalon and LCVR voltages, and the state parameter of the
phase diversity mechanism. Likewise, the accuracy in
the camera synchronism with LCVRs is better than one ms. The demodulation coefficients are
received from the CSW and then applied (if necessary) to the polarized images
making the intermediate calculations in signed 32-bit words in 2's complement.
After demodulation, images are truncated (bit shifted) by 15 bits plus those
specified (3) for the different observing modes. As commented on earlier, no need for 
on-board demodulation was present during the first flight.

\subsection{Data Flow}
\label{sec:dataflow}

\begin{figure}
\centering
\resizebox{0.4\hsize}{!}{\includegraphics[width=\textwidth]{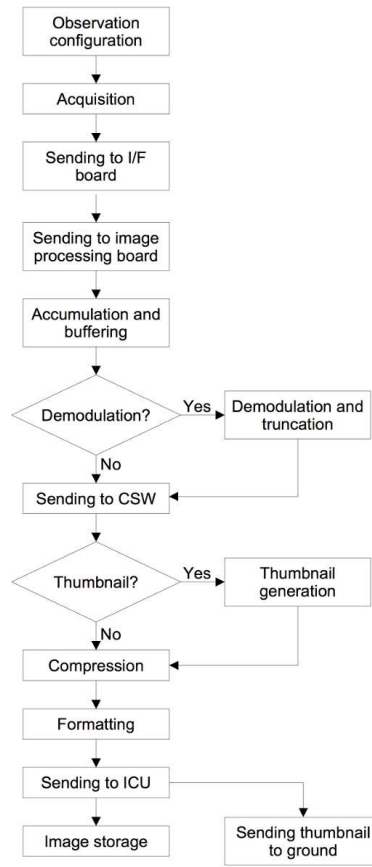}}
\caption{IMaX data flow.} 
\label{fig:data_flow} 
\end{figure}

In summary (see Figure\ \ref{fig:data_flow}), once the observing mode (see Section\
\ref{sec:optics}) is selected, the CSW sends the configuration parameters to
the DSW. The latter sends the corresponding commands to the cameras (exposure
time and mode) and to the PE (LCVR and etalon voltages plus the state of the
phase diversity mechanism). Images are acquired and then sent to the I/F board
that stores one of them per camera. The two images are then sent to the IPB for
them to be accumulated and buffered (according to the current observing mode).
After a complete set of images is obtained they are demodulated and truncated
(if needed) and sent to the CSW that is in charge of compressing (JPEG-LS) and
formatting them. They are finally sent to the ICU for storage. Truncation is
made by 3-bit shifting, hence removing the three less significant bits that are
enough to preserve a S/N of 10$^3$. \Sunrise telemetry
bandwidth is not wide enough for sending all the images to ground. The on-line
monitoring needs make it necessary to specifically process a sub-set of the
images for them to have a smaller size. These particular images are called
thumbnails and constructed by either binning or cropping the original ones
(that are nevertheless preserved) by a programmable factor. Thumbnails are sent
to the ground station by TMs.

\section{Thermal Design}
\label{sec:thermal}

The IMaX thermal design was conceptually envisaged to be as simple as possible while, at
the same time, capable of providing temperatures within the working range for
the various sub-systems. First of all, and in order to obtain an effective
thermal control, the IMaX OBE was conductively and radiatively isolated from
the environment. It uses a radiator as a top cover to reject the excess heat
from the dissipative elements.  Second, a sensible arrangement of the elements
on the optical bench was performed together with a careful thermo-mechanical
design including the selection of proper coatings and materials. The OBE is
attached to the Post-Focus Instrumentation platform (PFI) by kinematic
mountings at three articulate joints that provide a poor thermal conductance
(the three brown isostatic mounts in the bottom panel of Figure\
\ref{fig:optics3}). In order to maximize radiative isolation from the
environment, the lateral sides of the OBE (seen in the bottom image of Figure
\ref{fig:IMaX}) were finished in Alodine.
The bottom side was covered with a polished aluminium foil and the outer
radiator surface has been painted in white with Aeroglaze 9929 and Aeroglaze
A276 coatings.  The interior of the OBE was finished in black with a double
purpose, stray-light avoidance and better thermal coupling between
the elements.

From a thermal point of view, the OBE is divided into two areas. The area where
the pre-filter, the LCVRs and, the etalon are located uses active heating to provide a
thermally stable environment. The requirements for each optical sub-system have
already been mentioned, but the idea is that this area of IMaX has optical
components with temperatures in the range of 25 $^\circ$C to 35 $^\circ$C and with
active control implemented to various degrees of accuracy. The LCVRs and
pre-filter mountings are heated with an eight W thermostat whereas the 
optical mountings on both sides of the TCPE 
used a 3.5 W thermostat.  The TCPE itself included a 20 W thermostat.
The other area of the IMaX OBE contains the CCDs and
the PEE with electronic components with temperatures in the range -10 $^\circ$C
to 40 $^\circ$C. Most of the power dissipation takes place in this area and
four amonia heat pipes are used to direct this heat to the radiator on top of
the OBE. Two heat pipes are connected to the sides of the PEE and one heat pipe
to the front side of each of the CCD cameras. 

During the flight, neither the LCVRs and pre-filter, nor the doublets and
lenses reached their nominal temperatures (35$^\circ$C and 30$^\circ$C)
respectively.  Instead they worked typically at 31$^\circ$C and at 27$^\circ$C
with a diurnal fluctuation of $\pm 1^\circ$C. It is clear that the power of
these heaters was underestimated for an Arctic flight. In part, this was due to
the fact that the power consumption of the instruments was considerably smaller
than what was originally envisaged, so the instrument power dissipation and
mean temperatures were smaller than expected (both in the OBE and in the MEE).
The consequences of these lower temperatures had to be considered on the data
calibration of the instrument (see Sections\ \ref{sec:polcal} and
\ref{sec:speccal}).

The MEE was located on one of the \Sunrise racks attached to the gondola rear part 
so that its surrounding environment was quite different from that of the OBE.
The electronic boards were bolted to an aluminium grid-frame fixed to the 
chassis box's bottom face. A fan was located in the middle of the box in order to 
provide a homogeneous interior temperature, reducing hot spots with a proper air 
speed that also helped increasing low temperatures where necessary.  
Six temperature sensors, type AD590 FM, were glued both on the box chassis and 
on the key electronics components to monitor their temperatures. MEE temperatures during 
the flight were found to be in the cold side of the expected range, reaching in some parts 
temperatures below zero. Nevertheless, the established temperature range for
the electronic components, -20$^\circ$C to 40$^\circ$C was never exceeded, with the unit 
working properly at all times. 

\section{Pressurized Enclosures}
\label{sec:enclos}

\begin{figure}
\centering
\includegraphics[width=7cm]{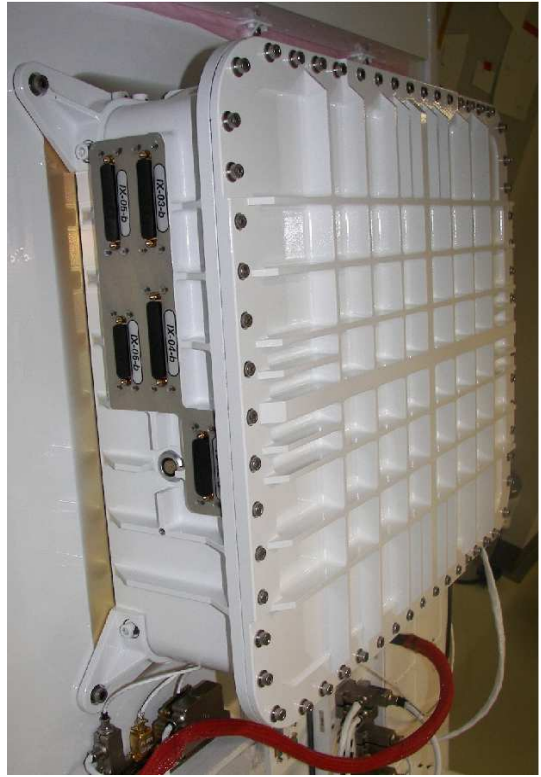}
\caption{The pressurized IMaX main electronics enclosure mounted into the flight panel.} 
\label{fig:MEE} 
\end{figure}

\begin{figure}
\centering
\includegraphics[width=9cm]{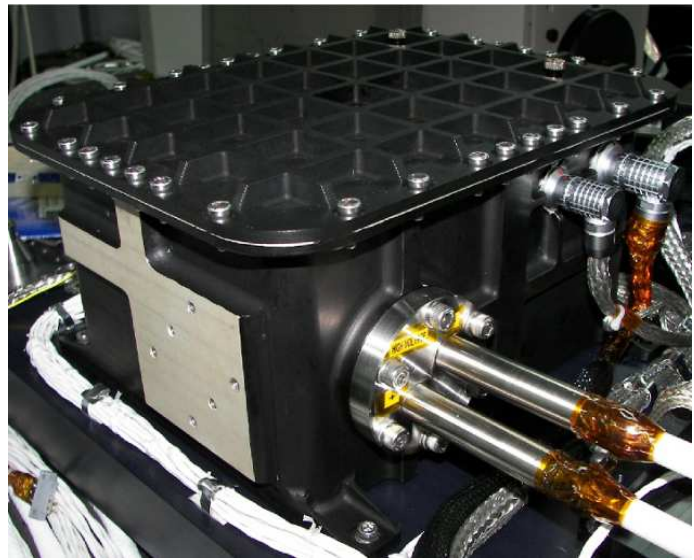}
\caption{The pressurized IMaX proximity electronics enclosure mounted in the optical bench.
The location of the lateral heat pipe (not shown) is visible. The high voltage 
connectors are seen in the lower right of the image.} 
\label{fig:PEE} 
\end{figure}

Two pressurized vessels have been developed to protect the IMaX electronics
(commercial grade electronics) from the mission environmental conditions: the
MEE and the PEE (see Figs.\ \ref{fig:MEE} and \ref{fig:PEE}). The MEE is
attached to the \Sunrise electronic platform and the PEE to the optical
bench. Since the optical elements must be extremely stable, the PEE design has
particularly addressed the stability of its interface with the optical bench,
while the MEE has avoided deformations in its interface with the \Sunrise
electronics platform during the mission.  Each enclosure (MEE and PEE) is
basically composed of a base, a top cover, and the electronics support frames.
The PEE, which contains the HVPS, also includes an internal insulator that
separates its interior into two compartments and that provides insulation between
the HVPS and the rest of the electronic components to avoid electromagnetic
interferences between them.  Following thermal design guidelines, the outer MEE
surfaces have a layer of thermal paint to optimize heat rejection over the
Alodine 1200S treatment, while the outer PEE
surfaces have been sulphuric anodized and black dyed like the rest of IMaX
opto-mechanical elements.

The base has an attachment flange to receive the seal gasket and the screws to
maintain, together with the top cover, the pressure at any stage of the
mission. Spring washers are used to maintain the sealing during flight time
and, additionally, all the screws are fixed with epoxy adhesive.  Strengtheners
that guarantee the stiff behaviour of the structure while maintaining a thin
wall (to reduce mass) have been included in both the base and the top cover.
Due to environmental temperatures and to external agents, resistance
fluorosilicone elastomer has been selected as an o-ring material for the
sealing.  A problematic area for any sealed compartment are the connectors. For
IMaX, we use hermetic connectors for data and power lines. A high voltage
connector (specified for up to 20 kV) has been included in the PEE for the HVPS.
Each enclosure includes two purging valves. This purging system allows the 
ambient air inside the box to be replaced with dry nitrogen once it has been
closed and sealed.  The enclosures were purged and filled with nitrogen at a
pressure of 1.2 atm.  No leakages were detected during the entire flight.


\section{Instrument Calibration}

This section describes the calibration of the three main functionalities
of the instrument (imaging, spectroscopy and polarimetry) in air and
in vacuum conditions as well as before and after integration with the
ISLiD and the telescope.

One of the major problems encountered during the AIV phase was the
reduced light levels with which the instrument was illuminated after
inserting the Barr pre-filter. Due to the small bandpass of this pre-filter
light levels were reduced dramatically even when high
power lamps were used. Often, the instrument had to use integration times of several
seconds for which it was not originally conceived.

   \begin{figure*}
   \centering
   \includegraphics[width=6cm,angle=90]{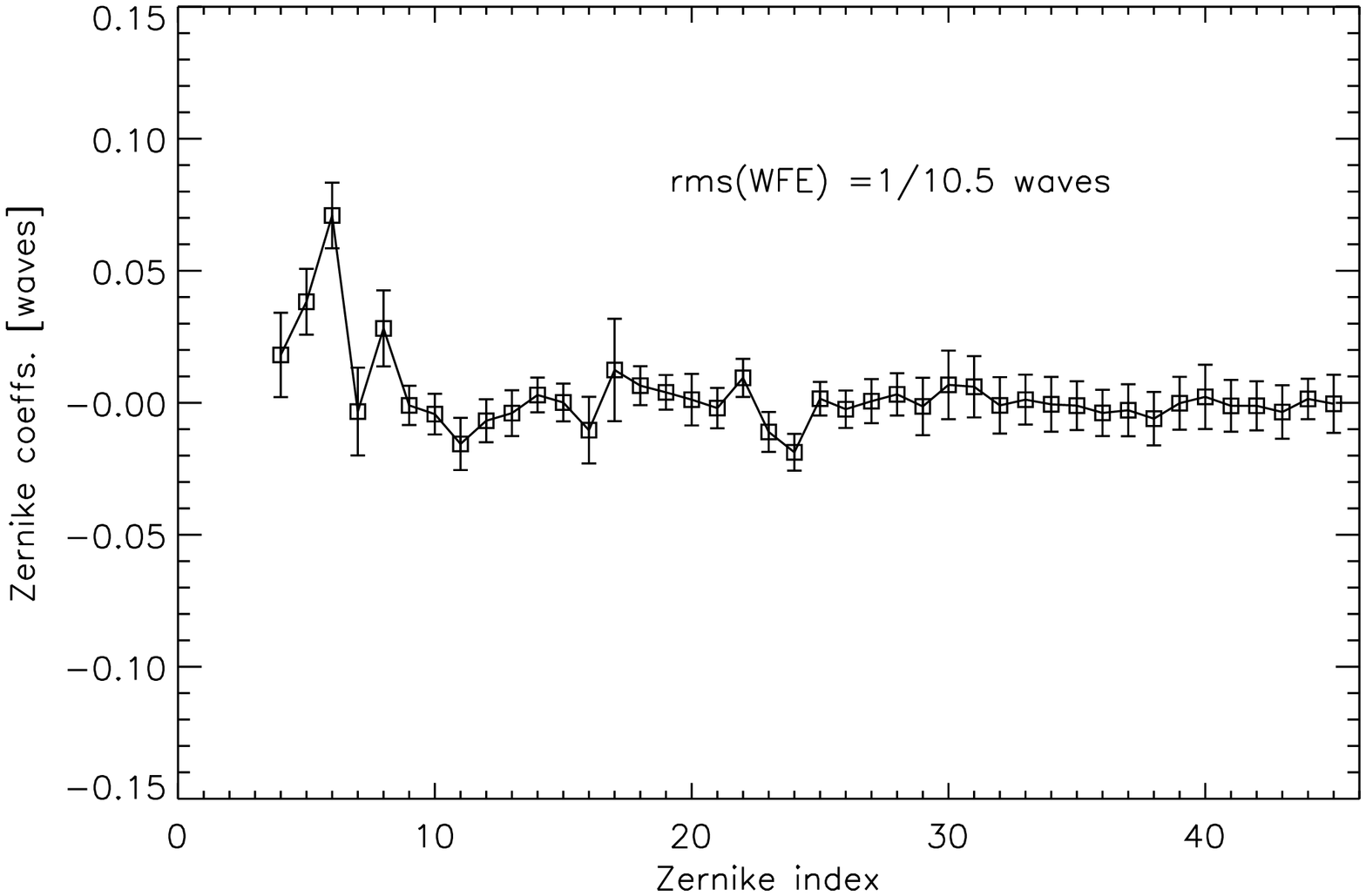}
   \includegraphics[width=6cm,angle=90]{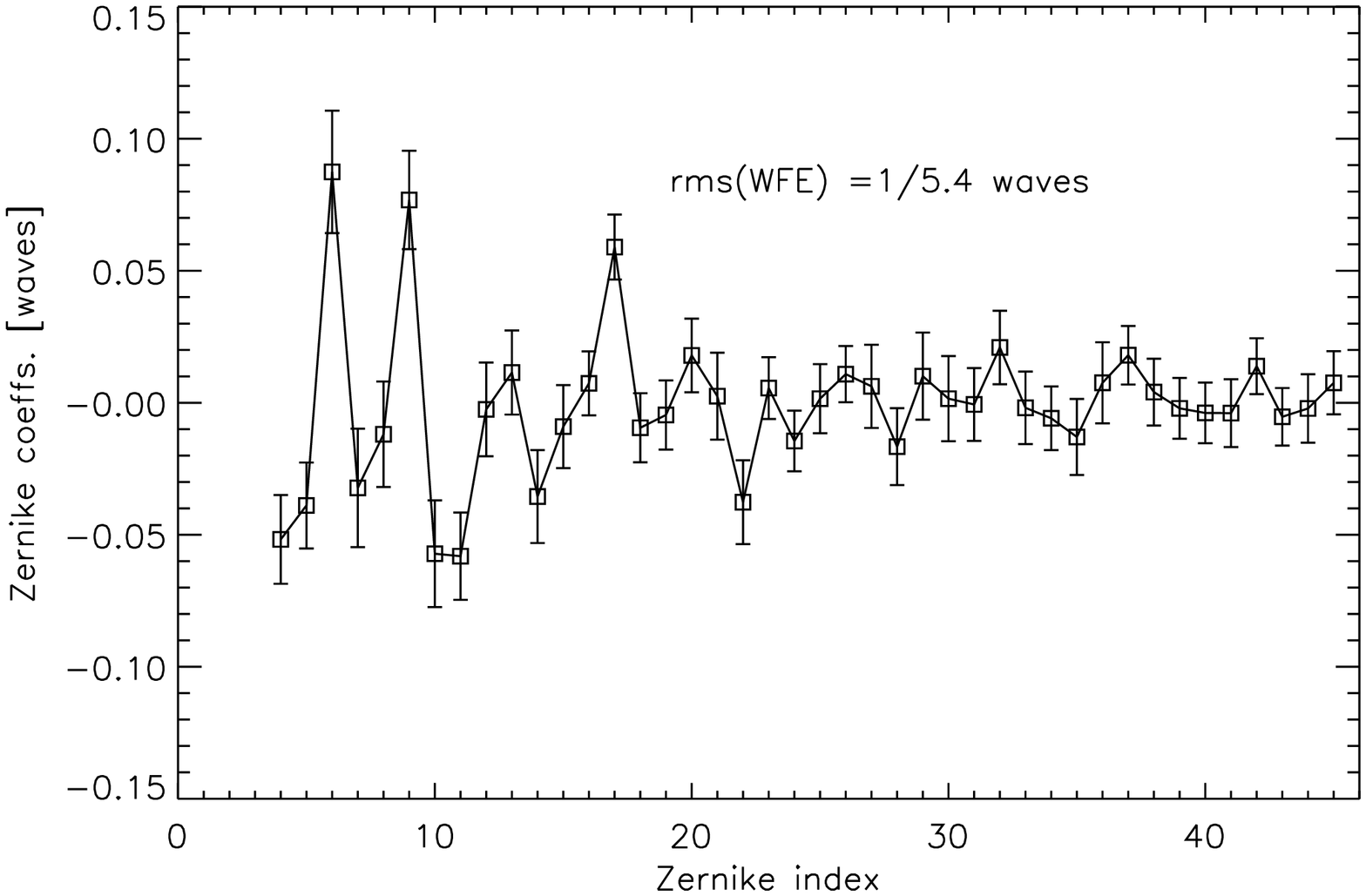}
      \caption{The Zernike coefficients describing the image performance of IMaX.
      Top: After integration at INTA with a global $\lambda$/10.5 rms WFE. The c6
      coefficient corresponds to the astigmatism generated by the etalon. Bottom:
      Performance inferred during the flight with a global rms WFE $\lambda$/5.4.
      A defocus component of 0.5 mm (at F4) is obtained.}
         \label{fig:imax_aberr}
   \end{figure*}

\subsection{Image Quality Characterization. Phase Diveristy}
\label{sec:iqpd}

Image characterization of the instrument at various stages during the
integration was done in two independent ways. First, standard interferometric
techniques at 633 nm were performed wherever possible. As the laser
light could not be used after insertion of the pre-filter, these measurements were
complemented with PD analysis to characterize the residual aberrations thanks
to  the built-in capability of the instrument. The use of the PD
technique has proven indeed very useful at all stages of the AIV phase.

After inserting all the elements with optical power, the optical axis of the
instrument was found with a two arcsec accuracy.  The telecentricity of the
instrument was checked by projecting an illuminated reticle at infinity (as
seen from F4) and visually observing the image of the reticle formed at the
pupil plane between the two folding mirrors behind the etalon. The
interferometric tests, using a retroreflecting ball system in each of the two
focal planes of IMaX, with the full optical setup of IMaX (without the
prefilter nor the etalon), showed the instrument optical quality to be
$\approx\lambda$/15, well within specifications. Before inserting the TCE-117
etalon, interferometric measurements were made to characterize its optical
quality. In spite of the fact that the manufacturer measured wavefront errors
were as small as $\lambda$/23 (see Table \ref{tab:acpoetalon}), we detected a
noticeable astigmatism component that degraded the quality to $\lambda$/11. The
reflection of a projected reticle by a theodolite on the etalon surface showed
an astigmatic figure that was incompatible with the manufacturer's
quotation.\footnote{We emphasize that the detected astigmatism was a figure
error due to a low order aberration. TCE-116 was, on the other hand, measured
to be at the manufacturer's value of $\lambda$/38.} After integrating the
etalon in the optical path, the combination of the $\lambda$/15 from the
optical components and the $\lambda$/11 from the etalon became dominated by the
latter.  Thus, the figure for the overall rms WFE of IMaX, as delivered for
integration in \Sunrise, was $\lambda$/11 (Strehl ratio of 0.7).  
This was confirmed by PD inversions made with a bunch of optical fibers 
located in F4  used as image object. 24 PD-pairs (at 522nm, the
pre-filter was not yet in) were acquired using the two cameras. Defocusing
here was performed by moving the fiber bunch through the focus position by known
amounts allowing the inference of the aberrations in both channels. The left panel of 
Figure\ \ref{fig:imax_aberr} shows the Zernike coefficients (from
the term four -defocus- onwards in the Noll, \citeyear{Noll76} basis), 
characterizing the aberrations of the system. 
The rms WFEs, derived from these calibrations (with the etalon
set to different voltages) range from $\lambda$/9 to $\lambda$/17 with a mean 
value of $\lambda$/11. In all cases,
the plots of the retrieved Zernike coefficients look very similar. A
common feature in all of them is a notable contribution of the 6th term (c6, third order
astigmatism) that should be ascribed to the
etalon. While the TCE-116 would have offered a better image quality
performance, the time needed to perfectly seal the TCPE of this etalon was
too long to consider its replacement and the decision to fly the TCE-117 was taken.

Once the pre-filter was inserted, which is close to a focal plane, these
calibrations were repeated with similar results, in particular 
the astigmatism represented by the 6th coefficient, but the data are noisier
due to the lower light levels. Additionally, the inclusion of the pre-filter
(that made the instrument monochromatic) brought up interference fringes
from various components that made the analysis more complicated (see Section 
\ref{sec:obsanl}).

The detected astigmatism implies the presence of a sagital and a tangential foci. The
distance between these two focus is 1.6 mm (at F4), a number similar to the focus 
depth. This implies that during integration of the instrument with the ISLiD, an
effort was needed to find the best compromise focus where the instrument performance
is maximized. This, of course, required aligning the two instruments below one mm 
along the optical axis. To this end, an illumination setup simulating light from the 
telescope and with an accessible pupil plane was established at MPS laboratories. 
A Hartmann test, masking the pupil in the horizontal and vertical direction, with IMaX at
various focus positions was made and the best sagital and tangential foci were established
to a fraction of a mm. Once at the air focus position, a PD calibration was made to
characterize the optical quality of the ISLiD+IMaX combination. Note that in this
case all optics is in place providing low illumination levels and generating interference
fringes. The etalon was tuned to maximum transmission 
of the pre-filter (voltages close to those used during the flight).
Yet the PD characterization made with a Siemens
star target at F2 (the ISLiD entrance focus)
resulted in a global WFE of $\lambda$/15 (and
with a similar c6 coefficient as observed with IMaX alone). This characterization
confirmed that we had found the best air focus, as no significant defocus term
appeared in the corresponding Zernike coefficient. This basically means  
that the ISLiD was transparent in terms of optical quality at the IMaX focal
plane. Clearly our portion of the ISLiD has an optical quality
well above $\lambda$/15. Once the air focus was found, IMaX was moved to the vacuum
position by displacing the instrument towards the ISLiD by 1.64 mm. Another
PD characterization was performed at this new position with the result that the
vacuum focus position was achieved, but the noise of the data was large enough not
to ensure a better focus position within one mm. The instrument was left in that position
for its transportation to ESRANGE.

All these calibrations provided a wealth of information that has been very
useful to understand the in-flight performance. The bottom
panel in Figure \ref{fig:imax_aberr}
provides a PD calibration performed the second day of the flight. Overall rms WFE
is $\lambda$/5.4 which is smaller than the performance 
we observed in the air focus of the ISLiD and IMaX combination. 
This discrepancy has to be understood in terms of the additional effects
introduced by the telescope and other in-flight generated aberrations. In
particular, the PD calibrations have shown the following effects to dominate:
\begin{itemize}
\item The PD check done at MPS on the translation of the air to vacuum focus was only 
able to confirm the vacuum focus position to within $\pm$0.5 mm. Values
near the extreme of this range can degrade significantly a $\lambda$/11 performance. 
Indeed, from the PD analysis of the
flight data shown in Figure\ \ref{fig:imax_aberr}, we deduced a defocus of a magnitude 
close to 0.5 mm.
\item The intrinsic astigmatism of IMaX (c6) is, of course, present but slightly
increased with respect to ground calibrations.
\item Trifoil terms (c9, c10) have appeared that we ascribe to the mirror mounting (but
have no real confirmation of this).
\item An spherical aberration is detected that was not seen before.
\end{itemize}

\begin{figure}
   \centering
   \includegraphics[width=7cm]{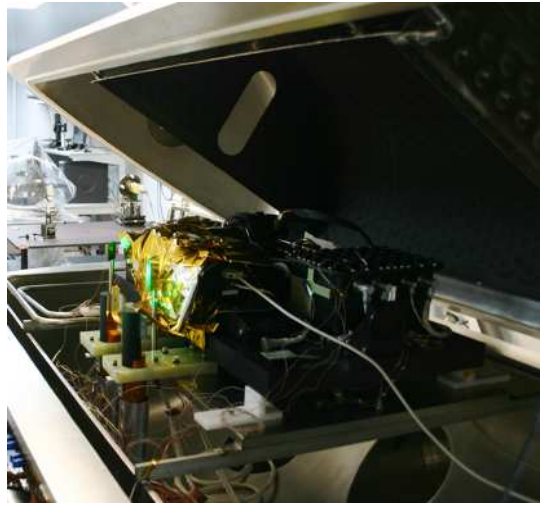}
      \caption{The IMaX optical unit in the thermal vacuum chamber at INTA
      facilities for testing the instrument performance under flight
      conditions. Light can be fed to the instrument 
      through an optical quality, flat window.}
         \label{fig:vacuum}
\end{figure}

Note, that we do not know what part of the defocus and spherical aberration
could originate in the lack of stable pointing of \Sunrise at the
level specified as explained in Berkefeld {\it et al.} (\citeyear{Berkefeld10}).  The
residual high frequency jitter, estimated from the Correlation Wavefront Sensor
(CWS) residual pointing errors, shows high frequencies ($>$30Hz) with
amplitudes of 0.03-0.04 arcsec (almost one IMaX pixel) whose direct effect
will be a blurring of the image similar to a defocus and/or a spherical
aberration.  As this jitter is associated with high frequencies one could
assume that this is a residual aberration that is constant in time much in the
same way as a real optical defocus of the system.

Despite the degraded performance as compared to IMaX on-ground calibrations,
having a WFE above the minimum threshold estimated by previous studies 
({\it cf.} Section\ \ref{sec:requirements}), makes us confident that near-diffraction
imaging is achievable with IMaX data (see Section \ref{sec:obsanl}).

\subsection{Vacuum Tests}
\label{sec:vacuum}

Vacuum tests of IMaX were included in the AIV plan with three interrelated
objectives: to test and verify the functionality at simulated (varying) flight
thermal and pressure conditions, to test the difference between the best focus
position in vacuum and in air, and to test the appearance or arcing in the
HVPS-etalon cabling system.

The IMaX optical bench was introduced in a thermal vacuum chamber that includes
optical quality flat windows that can be used to feed light into the instrument (see
Figure \ref{fig:vacuum}). The vacuum chamber has a base plate where the OB was mounted
and whose temperature can be controlled. 
The chamber also includes a shroud that surrounds the instrument
under test and whose temperature is also controllable. The temperatures
of these two elements were cycled between 16 and 24 $^\circ$C over several hours (to simulate the diurnal variation expected during the flight).

In order to check the image quality in vacuum, including checking the position
of the vacuum focus, a slit was placed at F4 and images were taken with the full 
system (the MEE was outside vacuum). This test was made without the pre-filter in the
instrument as this would have limited enormously the available amounts of photons. 
Instead a broad-band (100 \AA\ FWHM) pre-filter was used, thus allowing
slightly more than 50 orders of the etalon to contribute to the final image.
No heat pipes were still implemented in the instrument and heat from
the CCDs and PEE was evacuated with thermal straps to the baseplate.
The functional tests included LCVR modulation, application of  
etalon voltages, phase diversity mechanism insertion and image acquisition.
All heaters and thermal sensors were already integrated in the instrument
and their performance monitored during the test. In this case, all temperatures
achieved were nominal and no shortage of power was evidenced for the LCVRs, 
pre-filters, and etalon environment heaters. However, the temperatures in the
actual flight were cooler than the minimum one used in this test which
prevented them from reaching their expected values. All other instrument functionalities
tested in this vacuum tests resulted in nominal conditions with the exception
of the arcing observed in the connectors of the HVPS. When working
at three mbar and the applied voltage was higher than one kVolt (absolute value), 
arcing was observed between the inner and outer metallic components of the 
external connectors. To solve this problem, all metallic parts were removed from 
the external connectors and a ceramic solution was adopted.

\subsection{Polarimetric Calibration}
\label{sec:polcal}
   
   \begin{figure*}
   \centering
   \includegraphics[width=8.27cm]{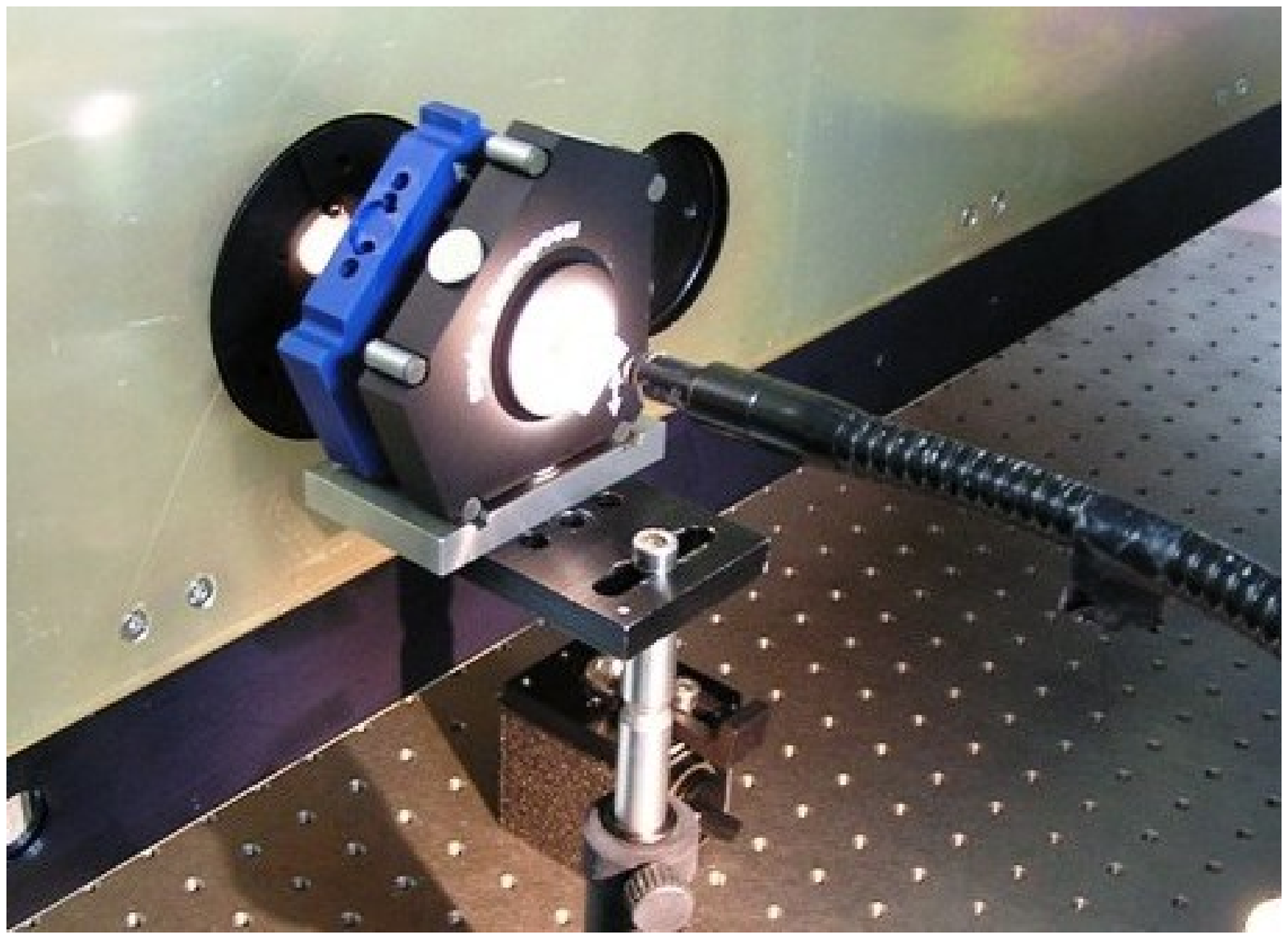}
   \includegraphics[width=8.27cm]{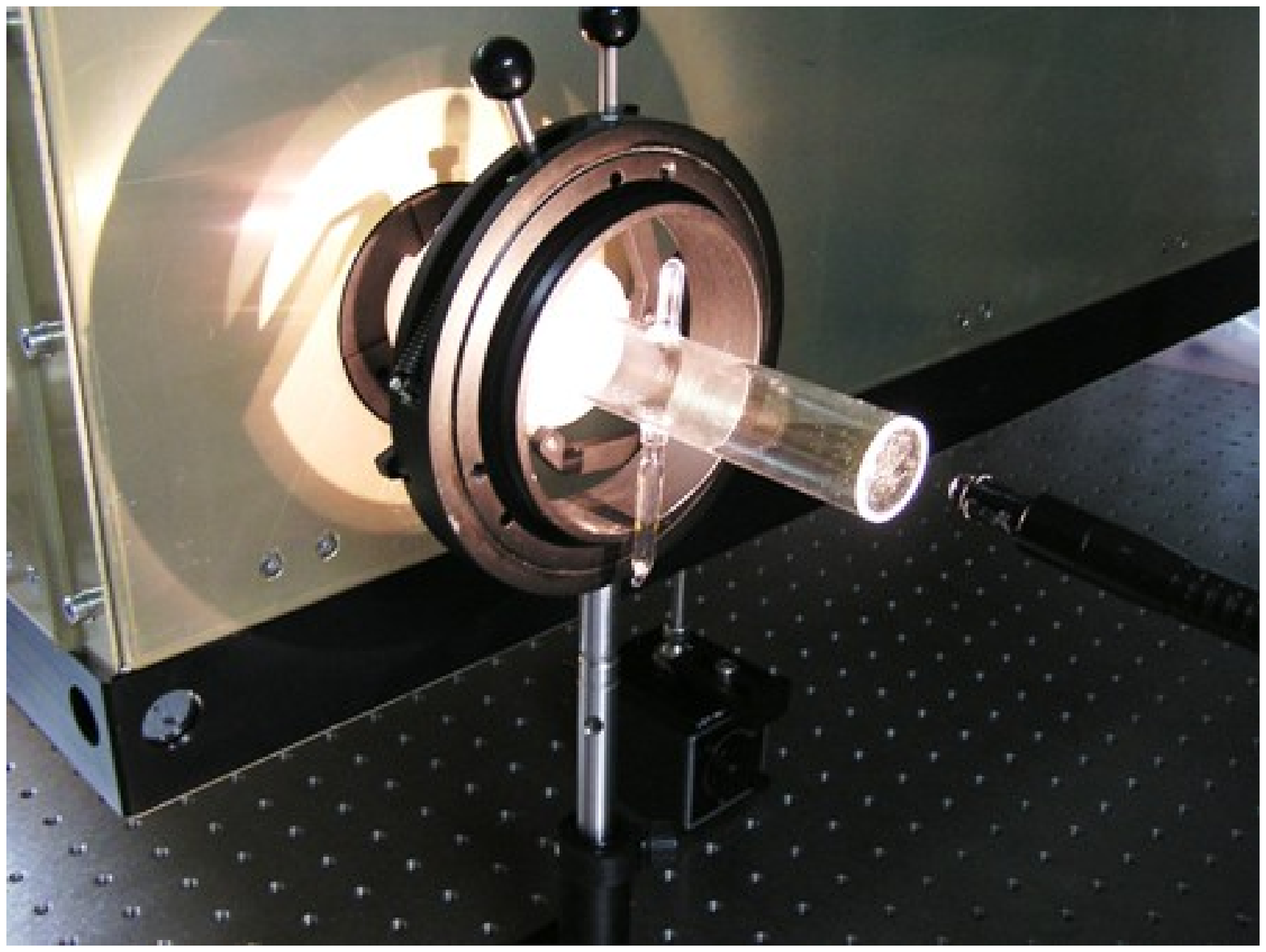}
      \caption{Top: IMaX polarimetric calibration with the eigenstate generator from 
      Meadowlark. Bottom: IMaX spectroscopic calibration using an iodine cell at the 
      entrance of the instrument.}
         \label{fig:cal}
   \end{figure*}

The polarization calibration of IMaX followed a two-step process. First, the
two LCVRs of the flight pair were characterized individually and then together
in a typical analyzer combination before integrating them into IMaX. 
The best voltage combination to maximize the
polarimetric efficiencies was found in this way (for the nominal flight
temperature, see Section  \ref{sec:lcvrs}). In this calibration, sufficient time
was given to the LCVRs to attain the corresponding retardance value and no
voltage optimization including finite switching time effects was performed.
The setup used in this calibration consisted of a fixed linear polarizer
followed by a quarter-wave plate\footnote{Whose exact retardance was found
by analyzing the results of a calibration run with no LCVRs inserted.}
in a rotating mount and stepped from 0 to 355
degrees every five degrees thus providing 72 calibration measurements.  Before the
detector (a photodiode in this case) a fixed linear polarizer was used as
analyzer (and oriented parallel to the entrance linear polarizer). This setup
allows a least-square determination of the ${\bf M}$ and ${\bf D}$ matrices and
their errors. The entrance linear polarizer was vertically oriented thus fixing the
X-axis (positive $Q$) direction of the modulation matrix.  Once the voltages
providing near maximum efficiencies were found, the two LCVRs were put into the
corresponding opto-mechanical mount and integrated into IMaX. A
characterization of the IMaX polarization modulation properties using the same
linear polarizer and rotating wave plate combination was then made. The
analyzer now was the instrument beamsplitter and the detectors, the instrument
CCDs. This calibration was made before inserting the pre-filter into the system
to maximize the light levels. A 100 \AA~pre-filter was used instead. Under this
configuration, two ${\bf D}$ matrices were derived using 1 and 0.25 s exposure
times and compared to investigate the effect of LCVRs finite switching times in
this exposing range. The differences between the various matrix terms were less
than 2 \% and no further optimization was considered necessary. The
efficiencies achieved under these conditions (broad-band illumination) were
$\epsilon_{Q,U,V}=0.56,0,57,0.52$, which are well above the requirements ({\it cf.}\ Section\
\ref{sec:requirements}).

After inserting the one \AA~FWHM IMaX pre-filter the usable light levels in the laboratory 
dropped dramatically as now only one etalon order enters the instrument. All 
calibrations were made with laboratory lamps and this filter required exposure times of one s 
and 20 accumulations at each polarization state in order to achieve decent signal levels. A
procedure including 72 measurements as before under these circumstances would have been 
prohibitively long. Instead, an eigenstate calibration system from Meadowlark Optics 
that generates, in a highly repeatable way, six polarization states was used (see Figure\ 
\ref{fig:cal}). This eigenstate generator allowed for orientation accuracies of 
one arcminute and the retardance of the quarter wave plate was known with
an accuracy of a tenth of a degree, ensuring calibration of IMaX 
demodulation matrices to better than 1\%.
The nominal linearly polarized states at 0$^\circ$, 90$^\circ$, 45$^\circ$ and 
135$^\circ$ together with the right and left circularly polarized states are generated to 
estimate the ${\bf M}$ and ${\bf D}$ matrices. Note that this time
24 measurements are used to evaluate the 16 matrix coefficients (instead of 288 as
before) and no practical inference of the uncertainties can be made. The only way to 
estimate the precision of the various matrices is by repeating the measurements
and observing the changes in their coefficients. 

This procedure was applied under three configurations: IMaX only, IMaX inserted in the ISLiD, 
and IMaX and ISLiD mounted on top of the telescope. In all cases, the vertical direction
was selected as the $+Q$ orientation. For the case of IMaX alone, the demodulation matrix 
obtained was (for camera 1 using vertical linear polarization):
\begin{equation}
{\bf D}_{\rm IMaX}=
\begin{pmatrix}
+0.25 & +0.25 & +0.25  & +0.25 \\
+0.41 & +0.41 & -0.43 & -0.36\\
+0.41 & -0.56  & -0.43 & +0.57\\
-0.58 & +0.59  & -0.39 & +0.40 
\end{pmatrix}
\end{equation}
with typical errors of $\pm 0.01$ in each of the coefficients (estimated from
calibrations made in different days and places). Camera 2 (horizontal linear
polarization) has a similar demodulation matrices with all signs in the
$D_{ij},~i=2,3,4$ terms exchanged.  The efficiencies in this case were
$\epsilon_{Q,U,V}=0.62,0,51,0.50$, again above the requirement level. After
insertion into the ISLiD and the telescope the measured demodulation matrix
was:
\begin{equation}
\label{eq:dsunrise}
{\bf D}_{Sunrise}=
\begin{pmatrix}
+0.25 & +0.25 & +0.25  & +0.25 \\
+0.45 & +0.37 & -0.37 & -0.36\\
+0.68 & -0.80  & +0.12 & -0.04\\
-0.07 & +0.17  & +0.58 & -0.68 
\end{pmatrix}
\end{equation}
with typical errors now of $\pm 0.02$ in each of the coefficients. The
corresponding efficiencies are: $\epsilon_{Q,U,V}=0.64,0,47,0.54$ still
fulfilling the original requirements but showing an imbalance between the $Q$
and $U$ efficiencies which translates into a S/N 1.4 times better in the
former Stokes parameter. The differences between ${\bf D}_{\rm IMaX}$ and ${\bf
D}_{Sunrise}$ can be ascribed to the folding mirrors present in the
telescope (M3 and M4) and the ISLiD (5 in total; see Gandorfer {\it et al.},
\citeyear{Gandorfer10}). Note that ${\bf D}_{Sunrise}$ shows that both
$I$ and $Q$ are evenly distributed among the four modulation states
$[I_1,I_2,I_3,I_4]$. However, $U$ is measured mostly in $[I_1,I_2]$ and $V$ in
$[I_3,I_4]$. As the number of accumulations is finally of the order of three to six,
this difference is not likely able to generate any systematic effect, but it is
convenient to keep it in mind.

The matrix ${\bf D}_{Sunrise}$ was measured at ESRANGE while the
LCVRs were at the nominal temperature of 35 $^\circ$C. As mentioned before
these nominal temperatures were never reached during science operations. The
LCVRs mounting showed a fluctuating temperature in the telemetry (according to
the daily variation) in the range $[29,31]~^\circ$C. In order to account for
this temperature difference, we have used the model mentioned in Section\
\ref{sec:lcvrs} to generate an additive correction to the involved Mueller
matrices and inferred a flight demodulation matrix of:
\begin{equation}
\label{eq:dflight}
{\bf D}_{\rm flight}=
\begin{pmatrix}
+0.25 & +0.24 & +0.26  & +0.25 \\
+0.46 & +0.36 & -0.38 & -0.34\\
+0.68 & -0.83  & +0.10 & +0.01\\
-0.02 & +0.14  & +0.57 & -0.69 
\end{pmatrix}
\end{equation}
with similar efficiencies. The differences between applying matrix
${\bf D}_{Sunrise}$ or ${\bf D}_{\rm flight}$ is that the estimated residual
cross-talk term $V\to U$ is reduced from 6 \% to 2 \%, respectively
(see Section  \ref{sec:datred}).

\subsection{Spectroscopic Calibration}
\label{sec:speccal}

   \begin{figure*}
   \centering
   \includegraphics[width=9.0cm,bb=30 0 420 280]{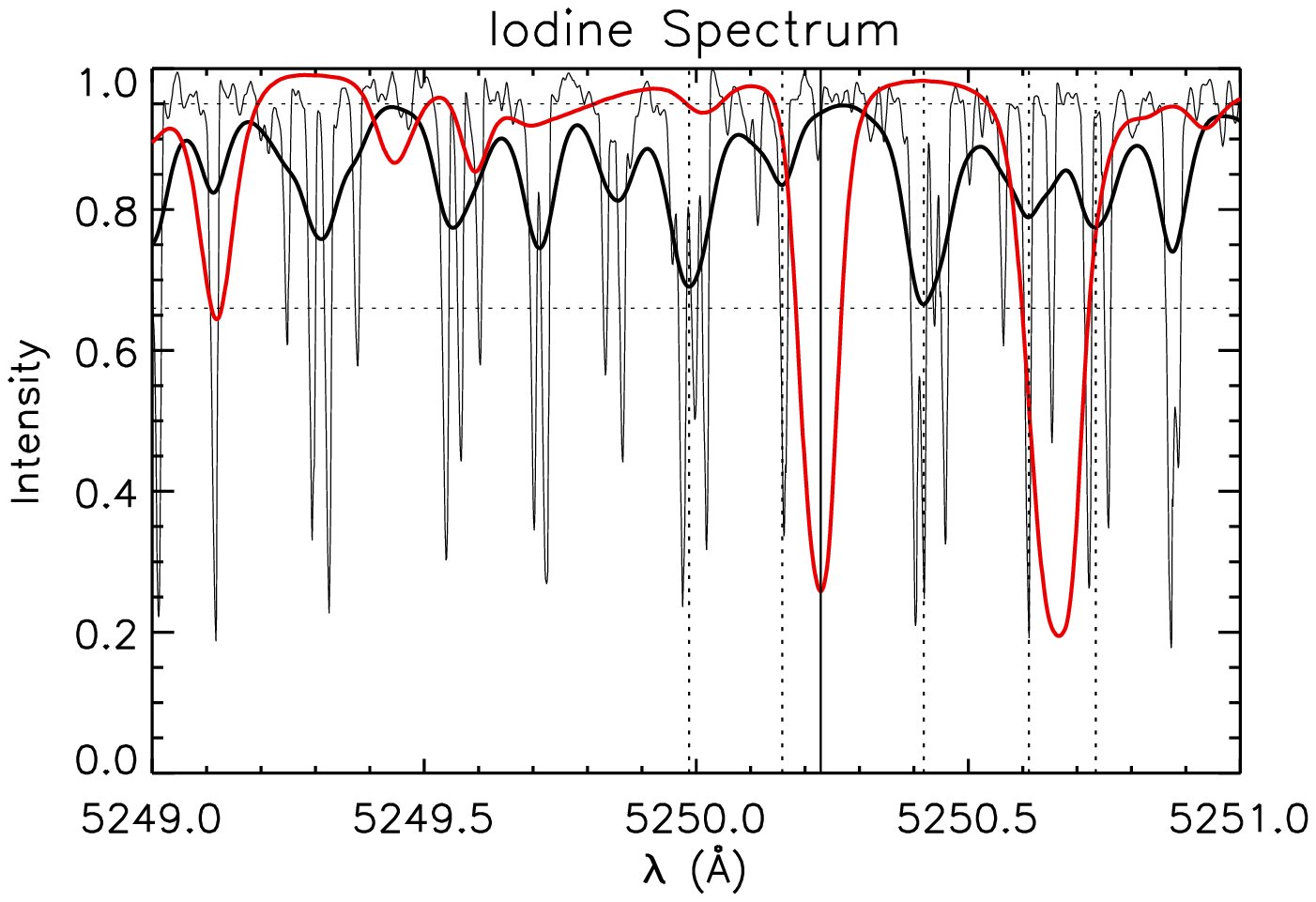}
   \includegraphics[width=9.9cm,bb=0 0 400 280]{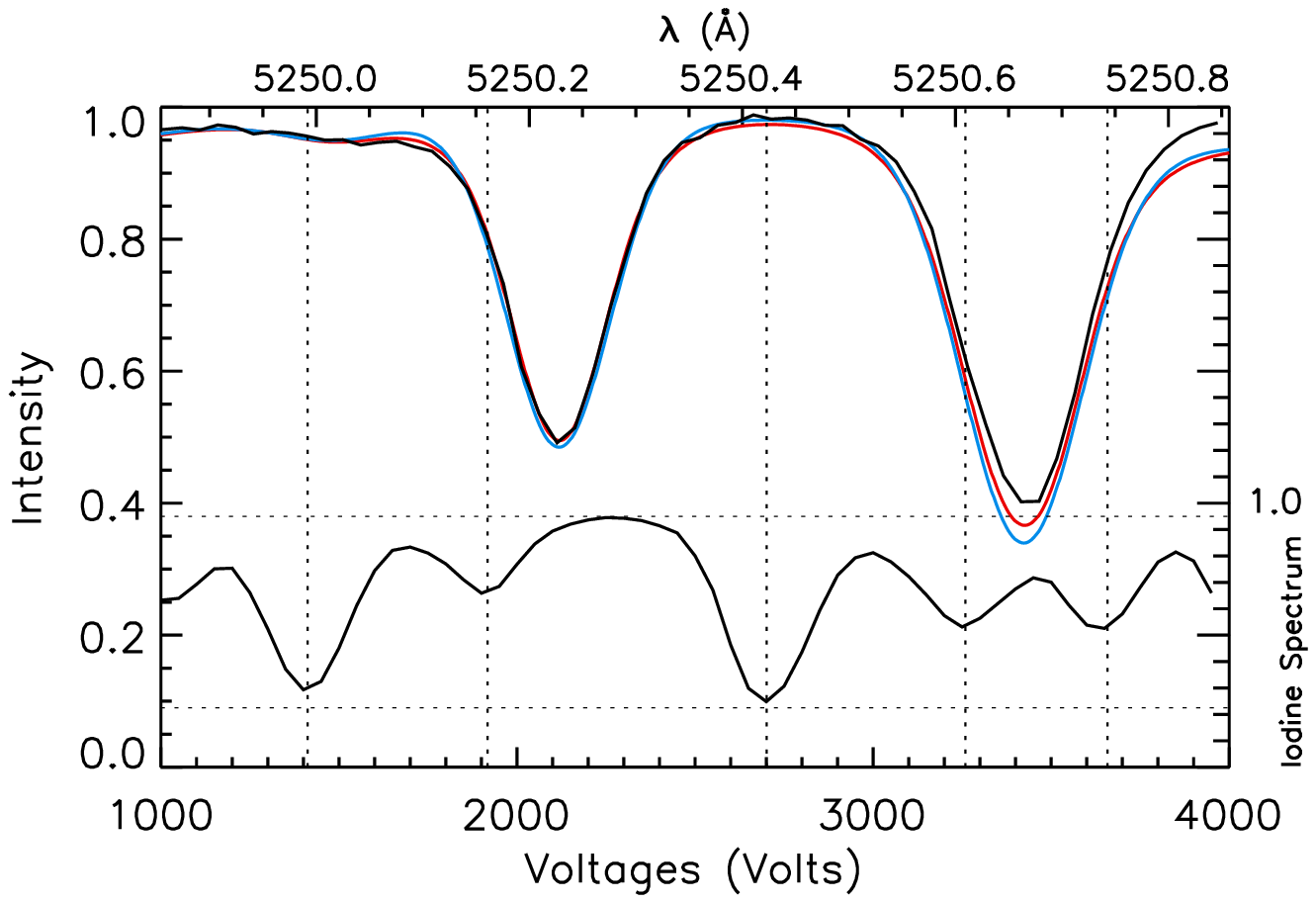}
      \caption{Top panel: The theoretical iodine spectrum is in thin solid
      line. The same spectrum after convolution with the IMaX theoretical
      profile is plotted in thick solid line. The red line represents the solar
      spectrum. The vertical solid line indicates the expected wavelength of
      the spectral line for the flight day. The dotted vertical lines mark the
      selected iodine blended spectral features used for wavelength
      calibration. Bottom panel: The black line on top is the solar spectrum as
      observed by IMaX at ESRANGE. The red line corresponds to the FTS spectrum
      convolved with the IMaX spectral point spread function. The blue line is
      the FTS spectrum convolved with an 85 m\AA~FWHM Gaussian. Bottom part
      (right axis label): observed iodine spectrum by IMaX during the
      calibration phase at INTA. This curve has to be compared with that predicted
      from the theoretical iodine spectrum shown by the thick solid line in the
      top panel. The vertical dotted line represents the
      blended iodine spectral features identified for the calibrations. The
      horizontal dotted lines are given as estimates of the detected intensity
      fluctuations.}
         \label{fig:iodine}
   \end{figure*}

The spectroscopic characterization of the instrument included two different
aspects. First, it had to provide the voltages under which the spectral line of
interest should be found.  Second, it should validate the instrument' spectral
function expected for an etalon with the properties given in Table
\ref{tab:acpoetalon} and thus establish the effective resolution.
As there was no option to feed the instrument with solar light during the AIV
phase at INTA, the decision was taken to illuminate the instrument using an
iodine (I$_2$) cell. Iodine cells have been extensively used for absolute
spectral calibration of spectrometers (Gerstenkorn and Luc,
\citeyear{Gerstenkorn78}). They provide a wealth of thin spectral lines that can be
used to establish a stable wavelength reference scale. Figure \ref{fig:cal}
(bottom panel) shows the iodine cell at the IMaX entrance during the AIV
calibration phase. The spectrum of I$_2$ in the wavelength region of interest
is shown by the thin solid line in the top panel of Figure \ref{fig:iodine}.
These lines have intrinsic widths of a few m\AA~that cannot be resolved by
IMaX. However, the convolution of the iodine spectrum with the Airy function of
the double pass etalon (see Figure \ref{fig:imaxpref}) shows a blended spectrum
with intensity fluctuations as large as 30\% (solid line in the same panel of
Figure\ \ref{fig:iodine}) which can be used to calibrate the instrument.  Also
shown is the solar spectrum (red line) given for reference. The vertical dotted
lines mark the five blended iodine lines selected for the voltage/wavelength
calibration. Applying voltages in the range of [0,4000] V and illuminating IMaX
with the iodine cell produced the spectrum shown in the bottom panel of Figure
\ref{fig:iodine} (bottom part; refer to the y-axis on the right). The
comparison of the voltages at which the reference blended lines of iodine are
found provides the voltage/wavelength calibration of the instrument:
\begin{equation}
\lambda-\lambda_0 =a+b\cdot V=-0.701+0.000335\cdot V
\end{equation}
with $\lambda_0 = 5250.225$ \AA; wavelengths are in \AA\ and voltages in V. The
$a$ coefficient gives the distance to the nominal wavelength of the line
($\lambda_0$) when the etalon is at 0 V. This distance to the spectral line of
interest changes depending on where on the solar surface the observations are
being made.  

Once the integration of the gondola, telescope, and PFI was completed,
\Sunrise was pointed to the Sun for a few hours (as allowed by the
weather conditions) and a basic calibration of the instrument could be
performed. In particular, light level tests and the first identification of the
spectral line were carried out. Figure \ref{fig:iodine} (bottom panel) shows the observed
spectrum that includes the two Fe {\sc i} lines at 5250.2 \AA~and 5250.6 \AA\ 
(solid black line). The top X-axis provides the wavelength calibration 
obtained for the applied voltages. In this case, the inferred $a$ coefficient was
-0.703 \AA~while $b$ stayed basically the same. The red line represents the
convolution of the FTS spectrum with the instrument profile from Figure
\ref{fig:imaxpref}, which produces a very satisfactory comparison. 
The spectral 
function used for the convolution has a range of four \AA. Using such a long
range can complicate the inversions of the
data. In order to provide a simpler description of the spectral properties
of IMaX, various Gaussians have been used to convolve the FTS data until a satisfactory
agreement with the observed spectrum in ESRANGE is obtained. The blue line corresponds to
a case where an 85 m\AA~FWHM Gaussian has been used. Thus, even if the FWHM
of the Airy function of the double etalon pass is of 65 m\AA, the effective resolution
of the instrument, including spectral stray light from the side lobes, is degraded 
to 85 m\AA.

While the voltages needed to map the IMaX spectral line according to this
calibration are in the range of [1800,2400]
V and of 2774 V for the continuum point (at +0.227 \AA~from line center),
the actual voltages used during the flight were higher as the ambient temperatures
never reached the expected values (those used for the calibrations). An automatic procedure
was implemented in the instrument software that applies voltages from 1000 to 3000 V every
100 V and automatically finds the voltage for the intensity minimum. A parabolic 
fit around this minimum value provides an updated $a$ coefficient.
The values of this coefficient obtained during the flight were around -0.783 \AA,
{\it i.e.} shifted, 80 m\AA\ bluewards. Given the temperature sensitivity
of this etalon (see Table \ref{tab:acpoetalon}), we deduce that it was
3.2 $^\circ$C cooler than when the calibrations were made on ground. This is a
reasonable number as the environment ``seen'' by the etalon was five degrees cooler, 
showing how effective the radiative coupling between the etalon and the surroundings
was in the instrument. The flight voltages were on average higher, being in the 
[2100,2700] V range for sampling the spectral line and of 3010 V for the continuum point.

The automated procedure was used every time an observing run was started allowing to 
identify the correct voltages to be used at the current \Sunrise pointing. Due to solar 
rotation the values of the $a$ coefficient varied in the range -0.7830$\pm$0.0325 \AA.

\subsection{Photon Budget}
\label{sec:photon}

The number of photons, $S$, reaching the IMaX detectors is given by
\begin{equation}
S=N_\odot \phi^2 \Delta\lambda\, A_{\rm D} \tau \, t_{\rm exp},
\end{equation}
where $N_\odot$ is the number of photons coming from the Sun at 525 nm per unit
time, area, solid angle, and wavelength interval.  The solar specific intensity
translates into $N_\odot=9.2~10^{17}$ photons s$^{-1}$~
cm$^{-2}$~sr$^{-1}$~\AA$^{-1}$.  $\phi^2$
represents the subtended angular area of the detector pixels on the plane of
the sky. $\Delta\lambda$ is the wavelength interval falling in one pixel and
can be assimilated to the spectral resolution of the instrument $\delta\lambda$
(85 m\AA). $A_{\rm D}$ is the effective collecting area of the
telescope and $t_{\rm ex}$ the exposure time of the individual
frames as given in Table \ref{tab:perf}.  
$\tau=\tau_{{\rm {\Sunrise}}}\tau_{\rm IMaX}$ stands for the
transmission of the complete system, by far the most uncertain factor in this
equation. While no precise measurements are available, estimates of the \Sunrise 
global transmission (including four telescope mirrors, four ISLiD mirrors, five
ISLiD lenses and three ISLiD wavelength dichroics) suggest $\tau_{{\rm {\Sunrise}}}=0.33$. 
The transmission of IMaX was also not accurately measured
because of the inherent difficulties of such a meaurement, but we estimate it
to be $\tau_{\rm IMaX}=0.13$. This transmission includes three mirrors, nine lenses, two
LCVRs, one pre-filter, one polarizing beamsplitter, and the four windows of the etalon
together with the two passes on the etalon itself. These two passes are assumed
to have a bandpass mismatch resulting in a transmission reduction of 70 \%
averaged over the entire surface. With these estimates, the global transmission
of the system is $\tau=0.043$.  The photon flux can thus be estimated to be
\begin{equation}
S=3.27\cdot 10^5 ~{\rm photons/pixel}.
\end{equation}
The detectors have a quantum 
efficiency of $Q=25$\%, which results in a number of detected electrons of
\begin{equation}
S_e=QS=8.1\cdot 10^4~{\rm e}^{-}{\rm /pixel}
\end{equation}
that fill 47 \% of the full well\footnote{This number has to be compared with
the 10$^7$ photoelectrons for one Gauss of sensitivity requirement introduced
early in the paper.}. This predicts a count level of 1925 counts per single exposure
(at continuum wavelengths) which is slightly higher than the observed count
level of 1500-1600 counts (DC offset and dark current amounts to 60 Digital
Units typically; see Table \ref{tab:ccdprop}). We ascribe this small
discrepancy to a slightly smaller overall transmission than the predicted value
($\tau$ of 0.034 instead of 0.043).  Photon noise and read-out noise add
quadratically to produce $\sigma_{\rm e}=289~{\rm e}^{-}$/pixel and, thus, the
signal-to-noise ratio of a single exposure was expected to be (for the
estimated transmission) $s/n=S_{\rm e}/\sigma_{\rm e}=280$. Now, using Equation\
(\ref{eq:numaccu}), one can see that, in order to achieve the final value of
$S/N\approx 10^3$ (700 per camera), we need to accumulate six frames in each of
the modulation states.  As the photon levels were slightly lower than
estimated, the S/N of the accumulated frames at continuum wavelength in IMaX
is indeed a bit lower than the expected 10$^3$ level. However values near 900
are routinely measured in the reduced data.

\begin{table*}
\caption{Observing modes of IMaX used during the 2009 flight} 
\label{tab:obsmodes} 
\centering 
\begin{tabular}{c c c c c c c c} 
\hline\hline 
Observing mode &  $N_\lambda$ & $N_{\rm p}$ & $N_{\rm A}$ & duration & S/N & Line samples & 
Continuum\\
 &   &  &  & $s$ &  & $({\rm m}\AA)$ & $({\rm m}\AA)$\\
\hline 
V5-6 & 5 & 4 & 6 & 33 & 1000 & -80,-40,+40,+80 & +227\\
V5-3 & 5 & 4 & 3 & 18 & 740 & -80,-40,+40,+80 & +227\\
V3-6 & 3 & 4 & 6 & 20 & 1000 & -60,+60 & +227\\
L3-2& 3 & 2 & 2 & 8 & 1000 &  -60,+60 & +227\\
L12-2& 12 & 2 & 2 & 31 & 1000 & -192.5 to +192.5 each 35 & +192.5\\
\hline 
\end{tabular}
\end{table*}

\section{Observing Modes and Data Analysis}
\label{sec:obmoddatan}

During the 2009 flight, \Sunrise observations in timeline
(automated) mode included a calibration run before and after the planned
scientific observations. These calibrations were done at disk center,
independently of the actual scientific pointing. They consisted of both
flat-fielding series, with the telescope moving in circles of 100 arcsec radius
around disk center, and imaging series for IMaX PD calibrations, with the
tracking re-enabled. IMaX was flat-fielded using the same observing mode
(see next section) as that used in the following scientific run. In each flat-field
series, 50 cycles of images (at each $N_\lambda$ position and $N_{\rm p}$ modulation
state) were acquired in timeline mode. After the scientific flat-fielding, the
etalon was set to the continuum wavelength point and 50 PD flat-field images
were recorded (with $N_{\rm A}=6$). Once the telescope was stopped and the CWS locked
at disk center, 30 frames (again, at continuum wavelengths only and with
$N_{\rm A}=6$) were acquired from which the system PSF can be inferred. IMaX dark
current frames were taken in accordance with the CWS calibrations that
commanded the dark current mask at F2.

\subsection{IMaX Observing Modes}
\label{sec:obsanl}

A number of observing modes are implemented in IMaX for its scientific
operations.
The modes differ in the number of wavelengths
observed, in the number of accumulations used, and in the number of
polarization states being obtained. The etalon is always set to a given
wavelength position, and the polarizations cycled $N_{\rm A}$ times. The total time of
one cycle in each observing mode is thus $N_{\rm A} N_{\rm p} N_\lambda t_{\rm ex}$, plus a
number of extra frames that are taken while the etalon is tuning to successive
wavelength points, which are discarded later.  As already commented before, and 
shown in Table \ref{tab:acpoetalon}, the tuning constant of LiNbO$_3$ etalons requires 
several hundredths of a milisecond in order to tune a typical wavelength jump of 40-80
m\AA. Thus, typically, IMaX discarded two or three frames every time the etalon was
being tuned to a consecutive wavelength, thus adding an amount of
$(2-3)N_\lambda t_{\rm ex}$ to the time given above. $N_{\rm A}$ is typically
targetted to the required S/N of 1000, although observing modes with smaller
signal-to-noise ratios, but offering faster cadences, are also implemented. The
number of polarizations is either four, for the full Stokes vector mode, or two for
the case of longitudinal observations only (including Stokes $I$ and $V$).
Table \ref{tab:obsmodes} lists all the observing modes used during the 2009
flight. The labels code the mode type (V for vector and L for longitudinal)
followed by $N_{\lambda}$ and, after a hyphen, $N_{A}$.  The most commonly used
observing mode was V5-6.  The continuum point used in most of the observing
modes (with the exception of L12-2) was set as the midpoint, in Stokes V
profiles of network points, between the 5250.2 and 5250.6 \AA~lines.

We note that longitudinal observing modes are implemented in IMaX as if they were
using four modulation states (like in vector modes). So L- modes are given in IMaX
data as groups of four sets corresponding to $I_{1,2,3,4}\propto [I+V,I-V,I+V,I-V]$. The 
demodulation matrices for the L- modes should then be prepared accordingly and have 
4$\times$4 dimensions (as for the V- modes).

\subsection{Data Reduction}
\label{sec:datred}
Before data reduction starts for a given observing run (identified
with an IMaX observing sequence number), several steps must be performed in advance.
The presence of interference fringes from some of the optical elements (the polarizing
beamsplitter and the Barr pre-filter) together with some dust particles in optical
elements near focal planes has considerably complicated the 
data reduction process. The presence of these unwanted patterns in the data was
anticipated, but the fact that they drifted slowly following the thermal (diurnal)
temperature changes of the instrument forces us to use always the closest flat-field
data set that is available. These diurnal temperature fluctuations were as large
as 3$^\circ$C as measured in some of the optical element mountings.
Additionally, the interference fringes do not remain
the same as the etalon is tuned to different voltages. While the exact reason
why this happens is unclear, we are certain that the etalon, which is close to a 
pupil plane, changes slightly the incidence angle of the rays and modifies the
fringe pattern for the various voltages used. The steps needed before a proper data reduction
can start are:
\begin{enumerate}
\item The closest flat-field run and PD set (including observations and PD flats) must be
identified. The closest dark frames 
must also be selected (although dark levels remained fairly constant throughout the flight).
\item A PD calibration consisting of inferring the instrument PSF as described 
by a set of 45 Zernike coefficients. A detailed description of the PD
inversion technique used here can be found in \citeyear{1996ApJ...466.1087P}
and references therein.
The retrieval of the system aberrations in every frame of
the calibration set (a total of the 30 realizations of PD-pairs) is performed
by applying PD-inversions independently in 10$\times$10 overlaped patches
of 128$\times$128 pixels. Since we have detected negligible anisoplanatism along the
whole field of view, the Zernike coefficients retrieved from all patches in all the
different realizations, are then averaged  to describe a mean wavefront and its
corresponding PSF (Vargas Dom\'\i nguez, \citeyear{Vargas09}).
\item Next, a valid mean dark current and flat-field frames are generated 
for each of the $N_\lambda\times N_{\rm p}$ states and for each camera
by simple averaging the individual realizations (typically 50).
\item The above generated flats are affected by intensity fluctuations produced
by the collimated configuration used for the etalon. These
fluctuations must not be included in the subsequent correction
of real data and must be removed ({\it i.e.}, accounting for the so-called 
blueshift effect in the flat-field data is necessary). The envisioned process leaves basically
untouched the intensity fluctuations due to interference fringes, dust particles, 
illumination gradients, {\it etc.}, so that they get corrected from the real data after
flat-field division. The main output from this procedure are the blueshift-corrected, 
flat-field  frames at each wavelength and polarization state.
This procedure also generates two additional important results. First, it calibrates
the $a$ coefficient (see Section\ \ref{sec:speccal}) that corresponds to the actual
pointing during the flat-field observations 
(in case they were not made exactly at disk center). Second, it generates a set of
coefficients for a surface fit to a third-degree polynomial that describes
the blueshift over the FOV. These
coefficients are later used for the pixel to pixel wavelength calibration. 
\item Once these flat-fields are produced, they are applied to the images of the
corresponding observing sequence that is farthest in time from the moment when
they were acquired (and that will be more affected by the thermal drift of
fringes and dust particles). The mean power spectrum of this sequence is used
to define a mask in the Fourier domain providing the frequency ranges where
residuals from the interference fringes are found. This mask will
be applied to the power spectrum of every single image in the observing
sequence, and the signal at the boundaries of the masked areas will be
interpolated so that a specific weighting function describing the excess of
power within these areas can be constructed. This function, specific for every
image, is finally used to filter out the excess of signal in the
real and imaginary parts separately.
\item The same set of images used to derive the mask above serves to generate an
additional mask (but now in the measurement domain) that marks the position of
dust particles in the frames. These positions will be replaced by
interpolating the signal at their boundaries, alleviating the effect of the
particles. Note again that the original flat-field is unable to fully account
for them due to their changing position as dictated by thermal drifts during
the flight. 
\end{enumerate}

Early in the process, when producing the averaged darks and uncorrected
flat-fields, the 936$\times$936 central pixels of every image are extracted in
order to eliminate the field stop used by the instrument at F4 as part of the
strategy to reduce instrument stray light. Therefore, all subsequent images have these
dimensions.
After all this preliminary information has been created for a given observing 
sequence (containing a number of cycles, each one with $N_\lambda\times N_{\rm p}$ images), the
data reduction process can proceed. The steps followed then, are: 
\begin{enumerate}
\item The $N_\lambda$ wavelengths and $N_{\rm p}$ states of a given cycle are dark
subtracted, flat-fielded with the images resulting from step four above, and
corrected for the pre-filter transmission curve (known from laboratory
calibrations).
\item The masks from steps five and six above are used to filter out the residual excess
power due to interference fringes in the Fourier domain and the dust particles residuals 
in the measurement domain.
\item The data are reconstructed using the PSF deduced from PD in step two above 
(whose Fourier transform is $S(${\boldmath $s$}$)$ with {\boldmath $s$} 
the 2D spatial frequency of modulus $s$) and a
modified Wiener optimum filter (from that originally
proposed in \citeyear{1971A&A....13..169B}). This
modification is implemented in order to avoid over-restoration in the high
frequency domain and introduces a regularization parameter $k$. 
Specifically, the reconstructed images are obtained after
computing the inverse Fourier transform of the product of the Fourier
transform of the image, $I(${\boldmath $s$}$)$, times $\Phi(${\boldmath $s$}$)$ defined as:
\begin{equation}
\Phi(\mbox{\boldmath $s$})
={{|I(\mbox{\boldmath $s$})|^2-|N(\mbox{\boldmath $s$})|^2}\over{|I(\mbox{\boldmath $s$})|^2}}
{{S^* (\mbox{\boldmath $s$})}\over{|S(\mbox{\boldmath $s$})|^2+
k\bigl (s/s_c \bigr )}}.
\end{equation}
$N(\mbox{\boldmath $s$})$ 
is the spectrum of the noise assumed here to be constant and evaluated
from the power observed near the cutoff frequency $s_c$. The regularization
parameter $k=0.005$ allowed to obtained contrast values of the restored
granulation similar to those obtained in ground-based short exposure images
while reducing the amplification of the noise at high frequencies.

The PD restoration technique is applied in the Fourier domain
that requires a FOV apodization that effectively reduces the IMaX FOV
in all of its sides by typically 30 pixels (thus leaving a usable FOV of
876$\times$876 pixels or 48$\times$48 arcsec).
The output from this step are both the non-reconstructed (Level 1) and 
reconstructed (Level 2) data.
Both data sets are treated similarly in the subsequent steps of the data reduction 
process. 
\item Data are demodulated using the ${\bf D}_{\rm flight}$ matrix.
\item The residual cross-talk with intensity is estimated at the continuum frame
and corrected for in all observed wavelengths. 
\item The intensities of the two cameras are equalized, the relative displacement between
the two found, and the data from the two cameras added to reduce jittering-induced cross-talk 
(which in this case corresponds, to first order, to some gradient along the direction
of the motion of the intensity frame 
that is transmitted to the polarization signals; see Lites \citeyear{Lites87}). Similarly,
the data from the two cameras are subtracted to produce the corresponding
jittering-induced polarization image.
\item The resulting frames from merging the two cameras are analyzed locally
(boxes of 52$\times$52 pixels or 2.86 arcsec square)
to ensure that global linear trends in the polarization observed in continuum
are not present and if so, removed. Similarly if local correlations with the 
jittering-induced polarization images produced in the previous step are found, they
are subtracted out.
\item Finally, residual cross-talk from Stokes $V$ to $Q$ and $U$ is searched
for (by correlating the corresponding frames) and if statistically significant
amounts of $V$ are found, a correction is applied. It is in this step where
using the matrix of Equation\ (\ref{eq:dflight}) instead of that in Equation\
(\ref{eq:dsunrise}) was noticeably better.
\end{enumerate}
The output of both Level 1 and Level 2 corrected data are finally written in FITS format.
\subsection{Sample Data and Calibration}
\label{sec:sample}

\begin{figure}
\centering
\includegraphics[width=0.6\textwidth]{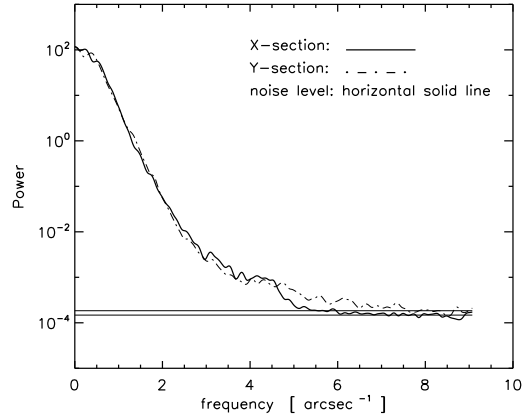}
      \caption{Power spectrum of an IMaX continuum image after PSF restoration in the 
      horizontal ($x$-section) and vertical direction ($y$-section). The horizontal
      lines represents the estimates of the noise level in the data.
	}
    \label{fig:power}
\end{figure}

IMaX data reduced as explained in the previous section correspond to the Stokes profiles 
sampled at various wavelength positions. At the time of writing this paper, 
this procedure has been
applied to several V5-6 (see Table \ref{tab:obsmodes}) runs obtained on 9 June 2009
(the launch was in the morning of the 8th). It has also been preliminary applied
to other data taken during the flight including the very last day, a few hours
before the balloon cut-off. These data indicate that the  
instrument performance was basically the same all along the flight without any noticeable
degradation. By selecting regions with no clear signs of polarization
signals on the reduced data, 
an estimate of the final S/N achieved in the individual wavelengths has
been made. For non-reconstructed data, values between 800-1100 are obtained and
the noise changes in random ways from pixel to pixel. We take this as evidence for
photon-noise dominated data. The situation for the PD reconstructed data is different, though.
The reconstruction process amplifies all frequencies but emphasizes more
the power at intermediate and high frequencies. This results in an increased noise level. 
Estimates made over the same regions as those made for non-reconstructed
data show noise levels a factor 2.5-3 times higher after the reconstruction process. 
Of course, the polarization signals themselves are also higher, but typically
by a factor 2, resulting in a polarization signal-to-noise ratio that is 
slightly smaller in the reconstructed data. But the S/N, 
that is typically referred to the continuum level, does not benefit from the increased
polarization signals and results in values of 300-400. Additionally, the noise
develops a spatial pattern of a few (2-3) pixels size that originates from the
reconstruction process at high frequencies. The free parameters (controlling the
enhancement of high frequency power) used in the reconstruction,
including the regularization parameter $k$,
were fine tuned to avoid clear signs of over-restoration in the polarization 
signals (as evidenced by the appearance of 
opposite polarity signals surrounding field concentrations), but still producing  
granular contrasts as found for the PD calibration data (which often reaches 14-15 \% at
disk center; non-reconstructed data show contrasts in the range of 8-8.5 \%. 

For the purpose of displaying the data and fully appreciate its quality, a
basic calibration has been performed. Stokes $I$ samples (4 inside the line
plus continuum) were fitted to a Gaussian providing the central wavelength, the
line center intensity, and the line FWHM. The central wavelength, after
correction for blueshift, provides the LOS Doppler velocity.  Longitudinal and
transverse magnetograms have been constructed by averaging the four samples
of $Q_i$, $U_i$ and $V_i$ inside the line. For Stokes $V$ the two red wavelength
points had their sign changed to avoid cancellation. In the case of $Q$ and
$U$, the average was done for $\sqrt{Q^2+U^2}$. Specifically, the following
samples were created from the data:
\begin{equation}
\label{equ:samples}
V_{\rm s}={{1}\over{4}}\sum_{i=1}^4 a_i V_i ~ ~\mbox{and}~~
L_{\rm s}={{1}\over{4}}\sum_{i=1}^4 \sqrt{Q_i^2+U_i^2},
\end{equation}
with the $\bar a$ vector being $[1,1,-1,-1]$. Note that the $V_{\rm s}$ and $L_{\rm s}$ samples have
half the noise of any individual wavelength. While these
samples provide polarization signals, they can be transformed into 
equivalent Gauss (or Mx cm$^{-2}$)
units in order to provide a better comparison with similar instruments. To this
end, the Fe {\sc i} line at 5250.2 has been synthesized in the 
\citeyear{1974SoPh...39...19H} quiet-Sun model using various
magnetic configurations. For the longitudinal calibration a set of field strengths
varying between 1 and 2500 Gauss aligned with the LOS was used. The same set of
field strengths is used for the transverse field calibration, but now with
an inclination of 90$^\circ$ with respect to the LOS and an azimuth
of 0$^\circ$. In both cases, the resulting profiles are convolved with
the IMaX spectral profile of Figure \ref{fig:imaxpref} and sampled
at the same four wavelengths as the real data. A linear least-square fit is
made in the range of [0,400] Gauss where the calibration curve still shows a clear
linear character, to provide\footnote{The minimum number of photoelectrons for 
a one Gauss detection in the
Introduction comes directly from the coefficients in Equation \ref{equ:magneto}.
}:
\begin{equation}
\label{equ:magneto}
B_{\rm L} (G)=4759{{V_{\rm s}}\over{I_{\rm c}}},~~\mbox{and}~~
B_{\rm T} (G)=2526\sqrt{{L_{\rm s}}\over{I_{\rm c}}},
\end{equation}
with $I_{\rm c}$ being the continuum intensity at the observed pixel. These equations allow
to transform (under the assumptions listed above) polarization fractions into equivalent
Gauss values. A 10$^{-3}$ noise level corresponds to a 4.8 Gauss longitudinal field
sensitivity and to an 80 Gauss transverse field sensitivity. These are typical sensitivities
of non-reconstructed magnetograms. For reconstructed 
data ($S/N \approx 350$), the sensitivities are 
14 Gauss and 135 Gauss respectively. But note that these
numbers heavily depend on the atmospheric model being assumed. This is 
particularly true for the extremely temperature sensitive line used
by IMaX. For example, if use is made of a network atmospheric model
(Solanki, \citeyear{Solanki87}), the constants 
in Equations (\ref{equ:magneto}) change to 11970 Gauss
and 3354 Gauss for the longitudinal and transverse cases, respectively.
Additionally, the line intensity minimum changes from 0.5 to 0.8, reflecting
the smaller absorption produced in a hotter atmosphere for this line.  Thus a
kG field produces a signal of 0.21 in $V_{\rm s}$ for the quiet-Sun model but of only
0.08 for the network one. For the present case, and due to the extremely quiet
nature of the observed regions, we consider more adequate to use the constants
in Equation (\ref{equ:magneto}) and keep in mind that they underestimate the equivalent fields
(or flux densities) by a factor 2, or more, in network concentrations. 
Note also that if one simply multiplies the $V_{\rm s}$ and $L_{\rm s}$ frames by these
constants, the effect of the blueshift of the line over the FOV is not
corrected for. In order to account for this effect, a calibration similar to
that described above, but with the samples shifted to the blue by known amounts
was made. This allows to understand how these constants change within the IMaX
FOV and produce a calibration frame (instead of a calibration constant) that when multiplied 
to the polarization samples, fully corrects the blueshift effect. The magnetograms presented 
in this section were transformed into equivalent Gauss values with these calibration 
frames and are, therefore, free from the blueshift over the FOV.

\begin{figure*}
\centering
\includegraphics[width=12.0cm]{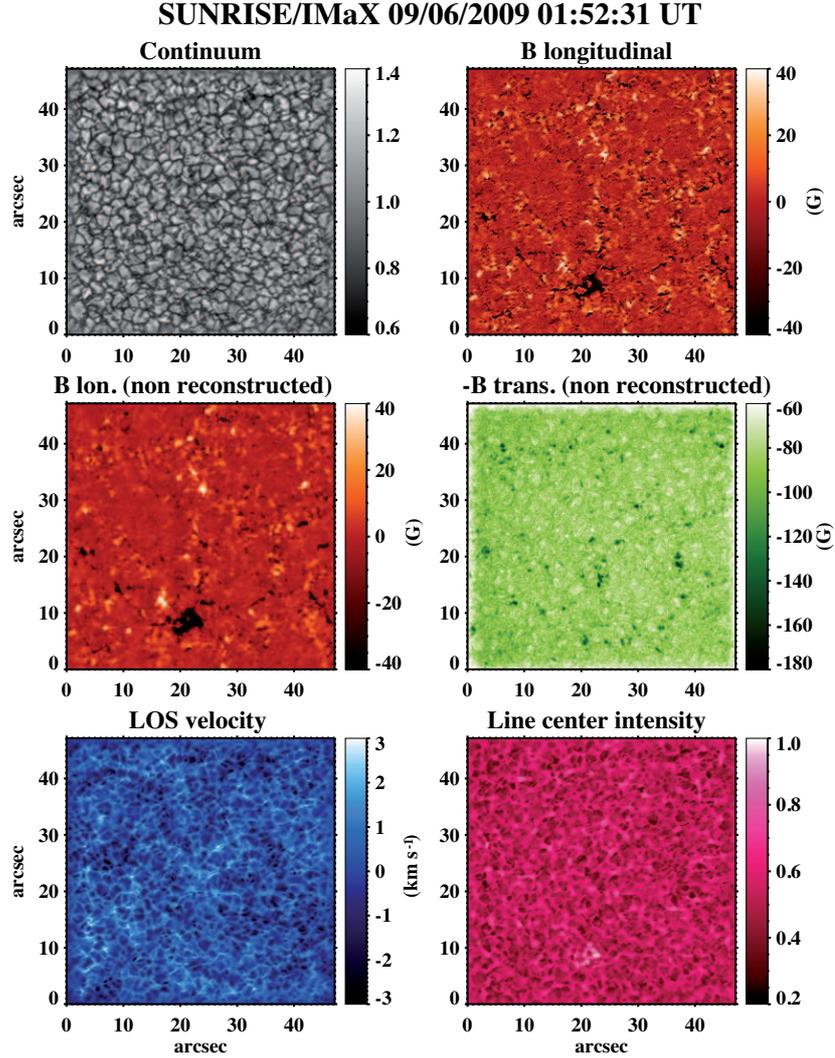}
      \caption{Sample data taken by \Sunrise/IMaX on the early morning of 9 June 2009, 
      a few hours after the launch. See text for details.
	}
    \label{fig:sample}
\end{figure*}

Figure \ref{fig:sample} shows an example of IMaX data after the reduction
process of section \ref{sec:datred} and calibrated as explained above. From top
to bottom and left to right, we show a continuum intensity frame, a
longitudinal magnetogram, the same longitudinal magnetogram with no
reconstruction process, the corresponding transverse magnetogram
(non-reconstructed), the LOS velocity frame and the line center intensity map. The
continuum intensity frame has an rms contrast of 12.5\%.  The effect of image
apodization can be seen near the boundaries of the image.  Two cuts of the
power spectrum of this frame in orthogonal directions are shown in Figure
\ref{fig:power}. If we estimate the noise level as marked by the horizontal
lines, the image power reaches the noise at frequencies corresponding to
spatial scales of 0.15-0.18 arcsec. This demonstrates the spatial resolution
reached by \Sunrise/IMaX. We stress that, by construction, the spatial
resolution is the same for continuum images, magnetograms, Dopplergrams, or for
any line parameters inferred from the spectral samples. The two longitudinal
magnetograms are provided to allow a comparison of the effects of image
reconstruction in them. The smaller sizes that are associated to the field
concentrations in the reconstructed data show the benefits of the process. A
more evident mixed polarity nature of the field distribution is also
noticeable. The downside is represented by the noise pattern that is stronger
in the reconstructed data. The scaling of the longitudinal fields ($\pm$ 40 Gauss)
corresponds to a polarization level $8\cdot 10^{-3}$ whereas the noise of this
magnetogram is $5\cdot 10^{-4}$ (non-reconstructed; for the reconstructed frame it is
$1.5\cdot 10^{-3}$) as benefited from the average over the four spectral points. The
transverse field image is represented in negative numbers to provide a better
visibility of the horizontal fields. This trick follows the work of Lites et
al. (\citeyear{Lites08}) who found internetwork horizontal signals similar to those
shown here. The equivalent Gauss levels found in this transverse map are about
180 Gauss before reconstruction and 260 Gauss after the reconstruction process. Given
the noise level of the latter data, the reconstructed transverse magnetograms
have a usability limited to the strongest concentrations. The transverse field
image shows evident signatures that resemble the solar granulation. While some
image jittering-induced cross-talk remains in the data, typically at the 1-2
$\cdot 10^{-3}$ level over intergranular lanes, the reason for being so prominent here
is different. Noise in the transverse data produces a positively defined veil of 80 Gauss 
from where real signals have to stand out. When no signals are present above this level,
the $1/\sqrt{I_{\rm c}}$ component of Equation (\ref{equ:magneto})
fluctuates according to the granular pattern and makes it this prominent.
The two last plots are derived from the Gaussian fits to the
Stokes $I$ profiles. The Doppler image has an rms velocity fluctuations of 0.73
km s$^{-1}$.  Extreme velocities (both upflows and downflows) have a magnitude
of 3.5 km s$^{-1}$.  The statistical errors (from the Gaussian fit) of the Doppler
velocities are in the range of 0.005-0.01 km s$^{-1}$ for the
non-reconstructed data and of 0.02-0.04 km s$^{-1}$ for the reconstructed ones. 
However, sometimes,
the simple Gaussian fit fails in the presence of strong network signals
as a result of the combination of the Zeeman broadening and the line weakening
due to the temperature excess. Under these circumstances, four
points cannot provide a good Gaussain fit (although velocities from an inversion code
that also uses Stokes $V$ are accurate).
The line centre intensity fluctuates in the range between
0.2, as found in small scale granular structures, to 0.9 in the strongest
network regions near the bottom of the frame. The mean value of this magnitude
is 0.5 as predicted by the Holweger-Mueller model and the IMaX spectral
resolution.  

\section{Conclusions}
\label{sec:conclu}

The IMaX instrument that flew in the \Sunrise stratospheric balloon in
June, 2009 for almost six days has been presented. The design, calibration, and
integration phases have been discussed in detail. No degradation of the
instrument performance was observed in the analyzed data so far. After its
recovery and translation to INTA facilities, the instrument was switched on and
images were acquired at nominal conditions with only some signs of optical
misalignment. The consumption of the instrument was at nominal values,
including the HVPS, showing that the LiNbO$_3$ etalon is acting as a capacitor
and suffered no structural damage. Although a careful spectral and polarimetric
end-to-end tests have not been performed yet, the instrument seems to have
survived the polar flight in rather good condition. 

The results presented here clearly show that IMaX/\Sunrise data will allow
solar magnetic and velocity fields tobe studied with a resolution of 0.15-0.18
arcsec for periods of times of around 30 minutes in the different observing
modes used by the instrument. Magnetic sensitivities are well below 10 Gauss for
longitudinal fields and 100 Gauss for transverse fields.  The \Sunrise
telescope was pointed to various places in the Sun, including a number of
intermediate positions between disk center and the solar limb.  These data sets
of unprecedented quality will provide a wealth of information about the 
quiet-Sun magnetic fields, dynamics, and interaction with the solar granulation.

\begin{acks}
The efforts put into this project by L. Jochum in its early phase
are gratefully acknowledged. IMaX was honored to have Juan Luis Medina
Trujillo as part of its team. He will remain forever in our memories.
The German contribution to \Sunrise is funded by the Bundesministerium
f\"{u}r Wirtschaft und Technologie through Deutsches Zentrum f\"{u}r Luft-
und Raumfahrt e.V. (DLR), Grant No. 50~OU~0401, and by the Innovationsfond of
the President of the Max Planck Society (MPG). The Spanish contribution has
been funded by the Spanish MICINN under projects ESP2006-13030-C06 and
AYA2009-14105-C06 (including European FEDER funds). The HAO contribution was
partly funded through NASA grant number NNX08AH38G.
\end{acks}

\end{article} 
\end{document}